\documentclass[preprint2]{aastex631}
\usepackage{gensymb}
\usepackage{natbib}
\usepackage{verbatim}
\usepackage{hyperref}
\usepackage{xcolor}

\usepackage{amsmath}

\newcommand{\uu}{\mathbf{u}}
\newcommand{\bb}{\mathbf{B}}
\newcommand{\avec}{\mathbf{A}}
\newcommand{\lnrho}{\ln \rho}
\newcommand{\cs}{c_\mathrm{s}}
\newcommand{\css}{c_\mathrm{s}^2}

\newcommand{\cm}{{\mathrm{\,cm}}}
\newcommand{\second}{{\mathrm{\,s}}}
\newcommand{\km}{{\mathrm{\,km}}}
\newcommand{\kms}{{\km\second^{-1}}}

\newcommand{\au}{{\mathrm{\,au}}}

\newcommand{\pc}{{\mathrm{\,pc}}}

\newcommand{\yr}{{\mathrm{\,yr}}}

\newcommand{\gm}{{\mathrm{\,g}}}
\newcommand{\gram}{{\gm}}

\newcommand{\msun}{{M_\sun}}

\newcommand{\K}{{\mathrm{\,K}}}

\newcommand{\muG}{{\mathrm{\,\ensuremath{\mu}G}}}

\shorttitle{Resistive Pseudodisks}
\shortauthors{V\"ais\"al\"a et al.}

\begin{document}

\title{Exploring the Formation of Resistive Pseudodisks with the GPU Code Astaroth}

\author[0000-0002-8782-4664]{Miikka S. V\"ais\"al\"a}
\affiliation{Academia Sinica, Institute of Astronomy and Astrophysics, Taipei, Taiwan}

\author[0000-0001-8385-9838]{Hsien Shang}
\affiliation{Academia Sinica, Institute of Astronomy and Astrophysics, Taipei, Taiwan}

\author[0000-0001-7706-6049]{Daniele Galli}
\affiliation{INAF--Osservatorio Astrofisico di Arcetri, Largo E. Fermi 5, I-50125 Firenze, Italy}

\author[0000-0002-2260-7677]{Susana Lizano}
\affiliation{{Instituto de Radioastronom{\'i}a y Astrof{\'i}sica, UNAM, Apartado
Postal 3-72, 58089 Morelia, Michoac\'an, M\'exico} }

\author[0000-0001-5557-5387]{Ruben Krasnopolsky}
\affiliation{Academia Sinica, Institute of Astronomy and Astrophysics, Taipei, Taiwan}

\email{mvaisala,shang@asiaa.sinica.edu.tw}

\begin{abstract}

Pseudodisks are dense structures formed perpendicular to the direction of the
magnetic field during the gravitational collapse of a molecular cloud core. 
Numerical simulations of the formation of
pseudodisks are usually computationally expensive with conventional CPU codes.
To demonstrate the proof-of-concept of a fast computing method for this
numerically costly problem, we explore the GPU-powered MHD code
\textit{Astaroth}, a 6th-order finite difference method with low adjustable
finite resistivity implemented with sink particles. The formation of
pseudodisks is physically and numerically robust and can be achieved with a
simple and clean setup for this newly adopted numerical approach for science
verification. 
The method's potential is illustrated by evidencing the dependence on the
initial magnetic field strength of specific physical features accompanying the
formation of pseudodisks, e.g. the occurrence of infall shocks and the variable
behavior of the mass and magnetic flux accreted on the central object. 
As a performance test, we measure both weak and strong scaling of our
implementation to find most efficient way to use the code on a multi-GPU
system. 
Once suitable physics and problem-specific implementations are realized, the
GPU-accelerated code is an efficient option for 3-D magnetized collapse
problems.  

\end{abstract}


\section{Introduction}

Pseudodisks are disklike density structures produced during the gravitational
collapse of magnetized cloud cores
\citet[][hereafter GS93I,II]{GalliShu1993I,GalliShu1993II}. Observationally,
they can be identified as flattened structures perpendicular to the magnetic
field orientation, with infall signatures on their radial velocity profile. 
In the direction perpendicular to the field, the motion is hindered relative to
that along the field. The Lorentz force deflects the infall motion from going
radially to the protostar, producing a density enhancement perpendicular to the
direction of the magnetic field lines. They are transitional, nonequilibrium
structures where the gas continues its collapse to the center. Infall motions
essentially dominate pseudodisks and are not rotationally supported as
Keplerian disks are. 

The predicted sizes of pseudodisks around young stars are on the order of
several thousand au.  One of the first of such observed objects was the large
flattened gas envelope of HL Tau, which was observed to have infalling motions
at a scale of $2000\au$ by \citet{Hayashi1993}. This source has also recently
been revealed by \citet{BroganALMA2015} in the inner $100\au$ radius, a thin
dust disk with concentric rings and gaps. The inner gas disk must be
rotationally supported, rather than being a pseudodisk like its large-scale gas
envelope. The Atacama Large Millimeter/submillimeter Array has shown several
other Class 0 sources with the characteristics of pseudodisks such as HH 211, HH
1448 IRS 2, L1157, B335, and NGC 1333 IRAS 4A (see Section
\ref{sec:observation}).

The theoretical concept of pseudodisks is rooted in an extension of the
self-similar singular isothermal sphere (SIS) collapse solution of
\citet{Shu1977} into a more realistic physical situation that includes the
effect of the magnetic field. Following \citet{TerebeyShuCassen1984}, who
introduced rotation into the SIS solution using a perturbative approach, GS93I
used a similar kind of perturbation analysis to account for the effects of
magnetic fields. The perturbation solution of GS93I includes the impact of
magnetic feedback on the gas dynamics. GS93II tested this perturbation solution
numerically. GS93I and II assumed an SIS threaded by a uniform magnetic field
as the initial state for simplicity. Later, \citet{LiShu1996} proposed
quasi-static magnetized toroids in force balance as initial states for the
gravitational collapse of magnetized clouds. 

Direct numerical MHD collapse simulations are computationally expensive, as
they in general require high-performance computing resources to be properly
executed for scientifically significant purposes. Magnetized collapse is
inherently a multiscale phenomenon, needing a large scale to start the collapse
and a small scale for the structures produced by the collapse (e.g., pseudodisk
thickness). That requires a large computational volume to be solved during the
time necessary for the collapse to occur; while that time is fortunately
usually not too long, in comparison to e.g. the lifetime of protoplanetary
disks, the computational time step (controlled by the Courant condition) is
often dominated in MHD problems by those computational cells naturally having a
locally large Alfv\'en speed (with a large field intensity but low $\rho$), and
usually leading to a time step much smaller than the collapse time scales. The
time step size may also be restricted by the need to simulate nonideal MHD and
other diffusive processes. 

We experiment with a new fast method with GPU technology to identify a
computationally cost effecient strategy for this problem. We explore the
characteristics of the pseudodisk structures formed during the collapse of a
nonrotating and uniformly magnetized centrally condensed cloud core using a
straightforward setup that includes a central gravity field. We vary the
magnetic field strength and resistivity and exploit the performance of GPUs
within numerically stable limits. Our simple approach serves as a proof of
concept, which is meant for further in-depth explorations when more physics is
implemented into the new higher-order code, {\it Astaroth}. This is the first
example of a magnetized collapse problem explored with such a high-order fast
code, where the artificial numerical code diffusivity can be significantly
reduced. The adopted parameters are exploratory for demonstrating such an
application with the Astaroth GPU code. Further explorations using more
realistic parameters and microphysics will be implemented in future work.

This paper is structured as follows. In Section \ref{sec:methods}, we describe
the computational methods, the relevant physical equations and parameters. In
Section \ref{sec:results}, we describe the results of physical modeling. 
In Section \ref{sec:code_test}, we show code result comparisons and GPU
performance measurements. In Section \ref{sec:disc} we discuss their
implications on the pseudodisk formation using this method, and summarize in
Section \ref{sec:conclusions}.

\section{Methodology} \label{sec:methods}

\subsection{Numerical Methods}

Astaroth \citep{Astaroth2020, Pekkila2019} is a library made for accelerating
high-order stencil computations, such as finite differences, using
GPUs.\footnote{Astaroth is open source and available freely under GPL3 license,
\url{https://bitbucket.org/jpekkila/astaroth/.}} Astaroth began as an
experimental, numerical code \citep{Astaroth2017,vaisala2017thesis}, designed
to accelerate the 6th-order finite difference method (FDM) as in the Pencil
Code \citep{pencilcode2021}. The currently available version of the Astaroth
code is more mature and flexible, which has been used for an MHD dynamo problem
\citep{Vaisala2021}. It supports a domain-specific language (DSL), which
describes the needed computational operations. Astaroth is structured to work
as an application programming interface (API), making it a relatively easy
add-on as an acceleration tool for an existing code, customizable to any
potential stencil computation. It supports node-level and multi-node
parallelization and has been tested to scale with 64 GPUs so far
\citep{Lappi2021, Vaisala2021,pekkila2021}. Astaroth is optimized for running
Nvidia CUDA-supported GPUs and AMD GPUs with AMD HIP interface, and the CUDA
devices were utilized in performing the current runs presented in this work.

The high-order FDM solver of Astaroth (a 6th-order FDM and a 3rd-order
2N-Runge-Kutta integration scheme) uses MHD equation in Pencil Code style
\citep{pencilcode2021}, with the novel addition of sink particles implemented
in this work. Equations are solved in a nonconservative form, and density is
expressed in a logarithmic manner. An isothermal equation of state is assumed.
Therefore, we need to solve for $\lnrho$, the velocity field $\uu$, and the
vector potential $\avec$. We express the continuity equation as 
\begin{equation}
\label{eq:continuity}
\frac{D \lnrho}{D t} = \nabla \cdot \uu  + \nabla_6(\lnrho, \uu) - \frac{S_M(\lnrho, \Delta t)}{\rho},
\end{equation}
where $D/Dt = \partial/\partial t + \uu \cdot \nabla$ is the advective derivative.
The $\nabla_6$ term represents the upwinding scheme of \citet{Dobler2006},
which prevents Gibbs phenomena like fluctuations, or wiggles, from emerging in
the density field. The $S_M$ term represents the density sink as explained in
Appendix \ref{sec:sink}. 
The momentum equation is expressed as 
\begin{equation}
\label{eq:momentum}
\frac{D \uu}{D t} = -\css \nabla\lnrho + \frac{\mathbf{j} \times \bb }{\rho} + \mathbf{g} 
\end{equation}
\begin{displaymath}
+ \nu [\nabla^2 \uu + \frac{1}{3}\nabla(\nabla \cdot \uu) + 2 \mathcal{S}\cdot\nabla\lnrho] 
\end{displaymath}
\begin{displaymath}
+ \zeta_\mathrm{shock}[(\nabla\cdot\uu)\nabla\lnrho + \nabla(\nabla \cdot \uu)] + (\nabla\cdot\uu) \nabla\zeta_\mathrm{shock}
\end{displaymath}
\begin{displaymath}
- \frac{S_M(\lnrho, \Delta t)}{\rho}\uu.
\end{displaymath}
In the magnetic force term, $\mathbf{j} = \nabla \times \bb = \nabla(\nabla \cdot \avec) - \nabla^2 \avec$ 
is the electrical current density, and $\bb = \nabla \times \avec$ is the magnetic field. 
The quantity $\nu$ represents the physical kinematic viscosity of the fluid,
set as constant, and $\mathcal{S}$ is the traceless rate of strain tensor. The
shock viscosity term containing $\zeta_\mathrm{shock}$ is an artificial
viscosity term and it is further discussed in Appendix \ref{sec:shockvisc}. The
gravitational acceleration is $\mathbf{g}$ as explained in Appendix
\ref{sec:sink}. The division by $\rho$ is in the sink term because we compute
the velocity field instead of momentum, as in \citet{Lee2014}.

The induction equation is expressed in terms of the vector potential $\avec$,
\begin{equation}
\frac{\partial \avec}{\partial t} = \uu \times \bb + \eta\nabla^2 \avec,
\end{equation}
using a diffusive gauge with a resistivity $\eta$. Here, while generally $\eta$
has the form of Ohmic resistivity, resistivity term should be seen as
representing the general diffusivity of the magnetic field. In the numerical
solver, all operations connected to the magnetic field use $\avec$. Any
expression of $\mathbf{j}$ and $\bb$ expressed above, is computed in terms of
$\avec$.

\subsection{Initial and boundary conditions}

For all simulations in this study, we used an evenly spaced 3-D Cartesian grid
with $N=256^3$ active grid points. This was the best compromise between the
available computing power, efficiency, and memory storage, with a first
exploratory experimental attempt using a full 3-D grid. This serves as a direct
benchmark for more complicated models to be explored in future works.

As an initial condition, we start with the density distribution of a SIS,
\begin{equation}
\label{eq:initrho}
\rho(r) = \frac{\css}{2 \pi G r^2},
\end{equation}
for $r > R_\mathrm{sink}$, where $R_\mathrm{sink} = 2\Delta x = 162\au$.  The
density is kept constant inside $R_\mathrm{sink}$ to avoid the singularity at
the center. Sound speed is set as constant $\cs = 0.35\kms$. 
The size of the whole domain is $20600 \au \times 20600 \au \times 20600 \au$
(or $ 0.1 \pc \times 0.1 \pc \times 0.1 \pc$),  with spatial resolution of
$\Delta x = 81\au$.  The latter is determined by both the chosen domain size
and the grid resolution. It is enough to sufficiently resolve the pseudodisk at
the scale that we are aiming in this study, making it the practical choice. The
initial mass at the center of the sink particle is set to $1.8\msun$, so that
it is sufficiently large to provide enough gravitational force to induce the
collapse of the density distribution. The high initial mass of the sink
particle is a consequence of the neglect of self-gravity in this model, as the
sink particle needs to be sufficiently massive to drive the collapse. The total
mass of the computational domain is $5.3~\msun$, including the sink particle
mass. The initial magnetic field is uniform in the vertical $z$-axis direction
with values as listed in Table~\ref{tab:models}.

The field $\lnrho$ is set to have continuous second derivative at all
boundaries, and the velocity boundary condition is set such that inflow from
the boundaries is prevented, but flow out of the boundaries is possible. The
magnetic field is kept fixed at the boundary. There is no initial rotation in
our models, as the focus of this work is on magnetic effects. 

\subsection{Parameter space}

The parameter space of the simulations is determined by three physical
quantities: the initial magnetic field $B_0$, the kinematic viscosity $\nu$ and
the resistivity $\eta$. Of these parameters, $B_0$ has values changing from
$30$ to $300\,\muG$, which covers all field strengths that can be expected in
star forming regions \citep[see, e.g.,][]{Tsukamoto2022}. Setting values for
$\nu$ and $\eta$ required a degree of testing and exploration.

A characteristic feature of a very high-order method, like the one we use, is
that numerical viscosity and resistivity are low \citep{axelnum}. Therefore,
all diffusive quantities have to be expressed explicitly, because the
resolution puts practical limits on how low the diffusive properties can be in
practice. This however has multiple advantages, an obvious one being that
getting a realistic estimate of the Reynolds numbers is straightforward.
Viscous and resistive Reynolds numbers are dimensionless quantities which
combine the characteristic velocities ($U$) and scales ($L$) of the system with
viscosity ($\eta$) and resistivity ($\eta$), namely $\mathrm{Re} = U\,L/\nu$
and $\mathrm{Re}_\mathrm{M} = U\,L/\eta$. Realistic values of Reynolds numbers
are orders of magnitude larger than any feasible numerical simulation
\citep[See e.g.][]{rincon2019}. Therefore, ideally one both aims to minimize
$\nu$ and $\eta$ within numerically stable limits and examine how their change
affects the simulation. 

Another benefit is to have viscous and resistive quantities to be explicitly
known and controlled. Conversely, the negative side of it is that their effects
have to be investigated. It is important to note that no numerical simulation
is totally free from these issues. In fact, lower order codes have numerical
diffusivities that are not explicit but are present due to the numerical method
use. \citet{McKee2020}, for example, estimated a numerical viscosity of the
order of $10^{23}\,\mathrm{cm}^2~\mathrm{s}^{-1}$ for their grid-based codes.
This is much larger than the values that our high-order code requires to be
stable. To estimate the effective Reynolds number of the simulation, such
effects have to be accounted for by performing e.g., dissipation tests, where
numerical results are compared to analytical/expected results, such as the
diffusion decay test in \citet{Astaroth2017} or the current sheet test in
\citet{Gardiner2005}.

To resolve a system with best accuracy, one aims to find the value of $\nu$
which is the smallest possible for the given resolution and additional physical
properties. We first identify the minimum values of $\nu$ and $\eta$ for which
the simulations worked without catastrophic numerical effects.
In this first application of {\it Astaroth}, we assume a constant value of
viscosity while for $\eta$ we adopt a simple scaling with density as
\begin{equation}
    \eta = \eta_\mathrm{c} \sqrt{\frac{\rho}{\rho_0}},
    \label{etasimple}
\end{equation}
where $\eta_\mathrm{c} = 10^{18}~\mbox{cm$^2$~s$^{-1}$}$   
and $\rho_0 = 4.1\times 10^{-20}\,\gram\,\cm^{-3}$ is the minimum initial
density in the box. This dependence is physically justified in the limit of low
ionization fraction $x_e$, where $\eta \propto x_e^{-1}$
\citep[see e.g.][]{Pinto2008} and $x_e\propto 1/\sqrt{\rho}$ for a simple
balance of cosmic-ray ionization and dissociative recombination.
The initial tests converged with the smallest stable value of the viscosity
$\nu_c = 10^{19}\,\mathrm{cm}^2~\mathrm{s}^{-1}$ and the smallest value of the
resistivity $\eta_\mathrm{c} = 10^{18}\,\mathrm{cm}^2~\mathrm{s}^{-1}$. Thus,
we set $\nu_c$ and $\eta_c$ for all models and vary the magnetic field's
magnitude as shown in Table \ref{tab:models}. We note that no direct
microphysics are adopted for the chosen $\nu_c$ and $\eta_c$, apart from
density dependence of the resistivity because these values are set by numerical
limitations in this work. However, in a real physical system, one possible
source for their enhancement can be diffusion caused by small-scale turbulence.
Nevertheless, estimating turbulent diffusion is nontrivial \citep[e.g.,
see][concerning the difficulties for estimating turbulence diffusion from
direct numerical simulation of turbulence]{kapyla2018}. 

\begin{deluxetable}{ll}
\tablecaption{List of models and their parameters.  
\label{tab:models}}
\tablewidth{0pt}
\tablehead{
\colhead{Model name} & \colhead{$B_0$} \\
 & \colhead{($\muG$)}
}
\startdata
B000 & $  0$ \\
B030 & $ 30$ \\
B060 & $ 60$ \\
B090 & $ 90$ \\
B120 & $120$ \\
B150 & $150$ \\
B180 & $180$ \\
B210 & $210$ \\
B240 & $240$ \\
B270 & $270$ \\
B300 & $300$ \\
\enddata
\end{deluxetable}

\subsection{Performance benchmark}

We measured the performance and scaling of the \textit{Astaroth} with the type
of setup intended in this work, using available local hardware.
\citet{Vaisala2021} and \citet{pekkila2021} have reported benchmarks for the
small-scale dynamo problem and the general performance. We use 4 Nvidia Tesla
P100 GPUs per node, with 16 GB per GPU, on the local ASIAA/TIARA servers. We
also use up to 32 Nvidia Tesla V100 GPUs available in an 8 GPUs per node
configuration, with 32GB per GPU in the Taiwan Computing Cloud (TWCC) of the
National Center for High-performance Computing (NCHC). \textit{Astaroth} was
compiled without GPUDirect remote direct memory access (RDMA) when using these
two clusters. GPUDirect RDMA allows for direct device-to-device communication
without need to pass data explicitly through the host. Based on the benchmarks
of \citet{pekkila2021}, it provides a small improvement on the performance.
However, GPUDirect RDMA was not, based on our testing, supported on the systems
that we used for our benchmarks.

We measured strong and weak scaling on multiple resolutions, and GPU counts by
measuring the time taken by a computational loop without extra diagnostics, by
running the simulation several hundred steps and taking the average of the
single step measurements. As the shock viscosity was judged to take a
particularly high cost on the performance, we measured the benchmarks both with
and without shock viscosity. The benchmark times are illustrated and analyzed
in Section \ref{sec:speedup}. 

This is not the only performance scaling benchmark. More general benchmarks of
\textit{Astaroth} have been published before in \citet{Vaisala2021} and
\citet{pekkila2021}, who made a robust comparison against the CPU performance.
In the benchmark in this study, we are primarily interested in the effects of
added features, such as the shock viscosity, in comparison with the more basic
version of \textit{Astaroth} explored in the earlier work.

\section{Results} 
\label{sec:results} 

Our numerical results are one of the first experimental benchmarks for a
6th-order FDM applied to the magnetized collapse problem with a central source
of gravity. We follow the dynamical evolution of the systems, and observe the
flattening of the density structure in pseudodisks, as depicted by isodensity
contours. The Lorentz forces slow the advection of the flow across magnetic
field lines. The resistive properties of the gas affect how close the field
follows the flow, and lead to the variation of the mass-to-flux ratio.

\subsection{Time Evolution} 

\begin{figure*}[htb!]
\gridline{\fig{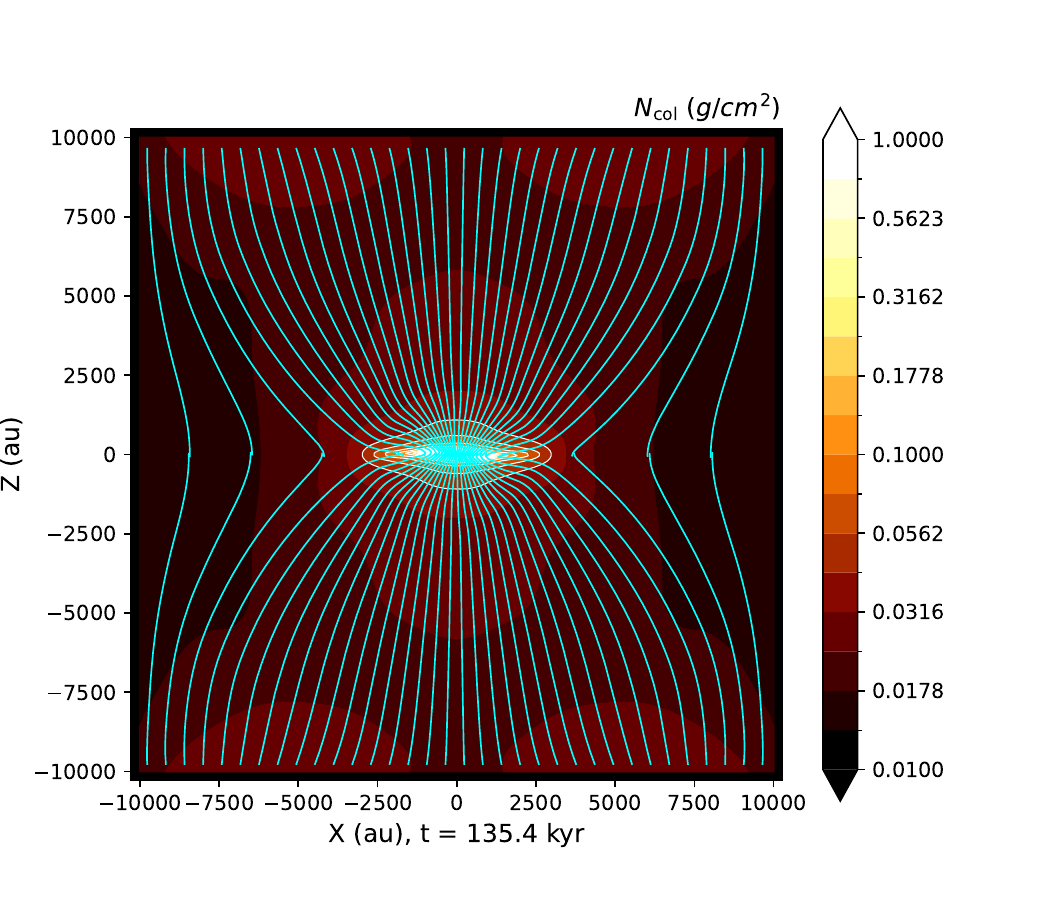}{0.5\textwidth}{(a) B030, $B_0 = 30 \muG$}
          \fig{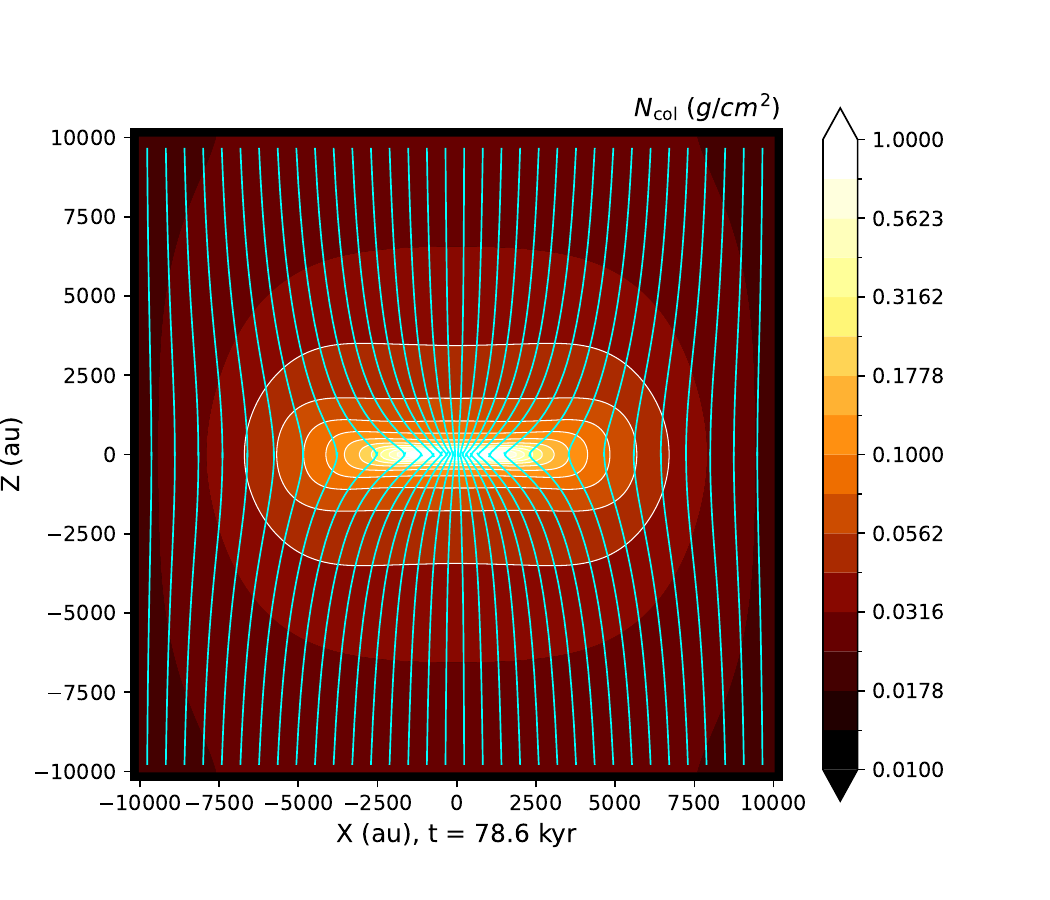}{0.5\textwidth}{(b) B150, $B_0 = 150 \muG$}
          }
\gridline{\fig{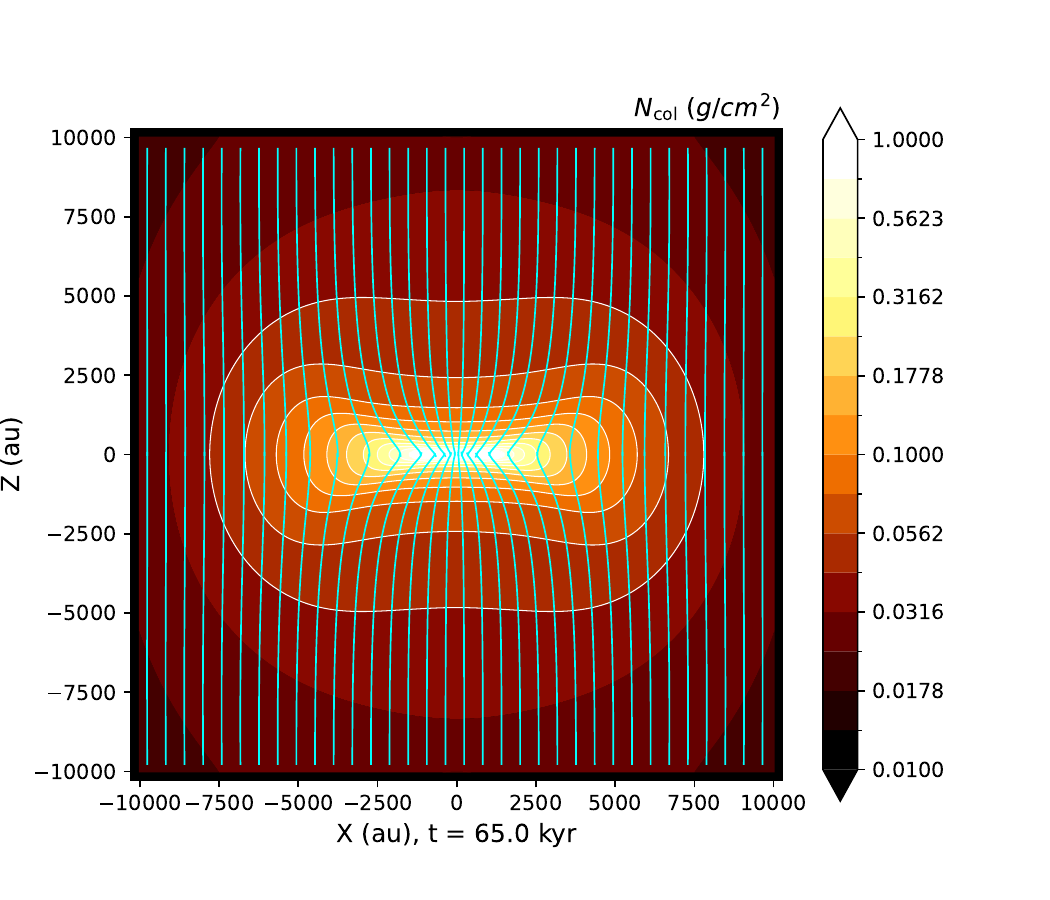}{0.5\textwidth}{(c) B270, $B_0 = 270 \muG$}
         }
\caption{Typical pseudodisk structures: (a) \textit{Weak field}, 
(b) \textit{Medium field}, and (c) \textit{Strong field}.
The color contours show column densities and the cyan lines trace the magnetic field lines on the XZ plane. 
The white contours follow the column density contours from the threshold of $0.0422\,\gram\cm^{-2}$ and above.
}
\label{fig:types}
\end{figure*}

\begin{figure*}[htb!]
\begin{center}
\includegraphics[width=0.8\textwidth, trim={0 3.3cm 0 4cm},clip]{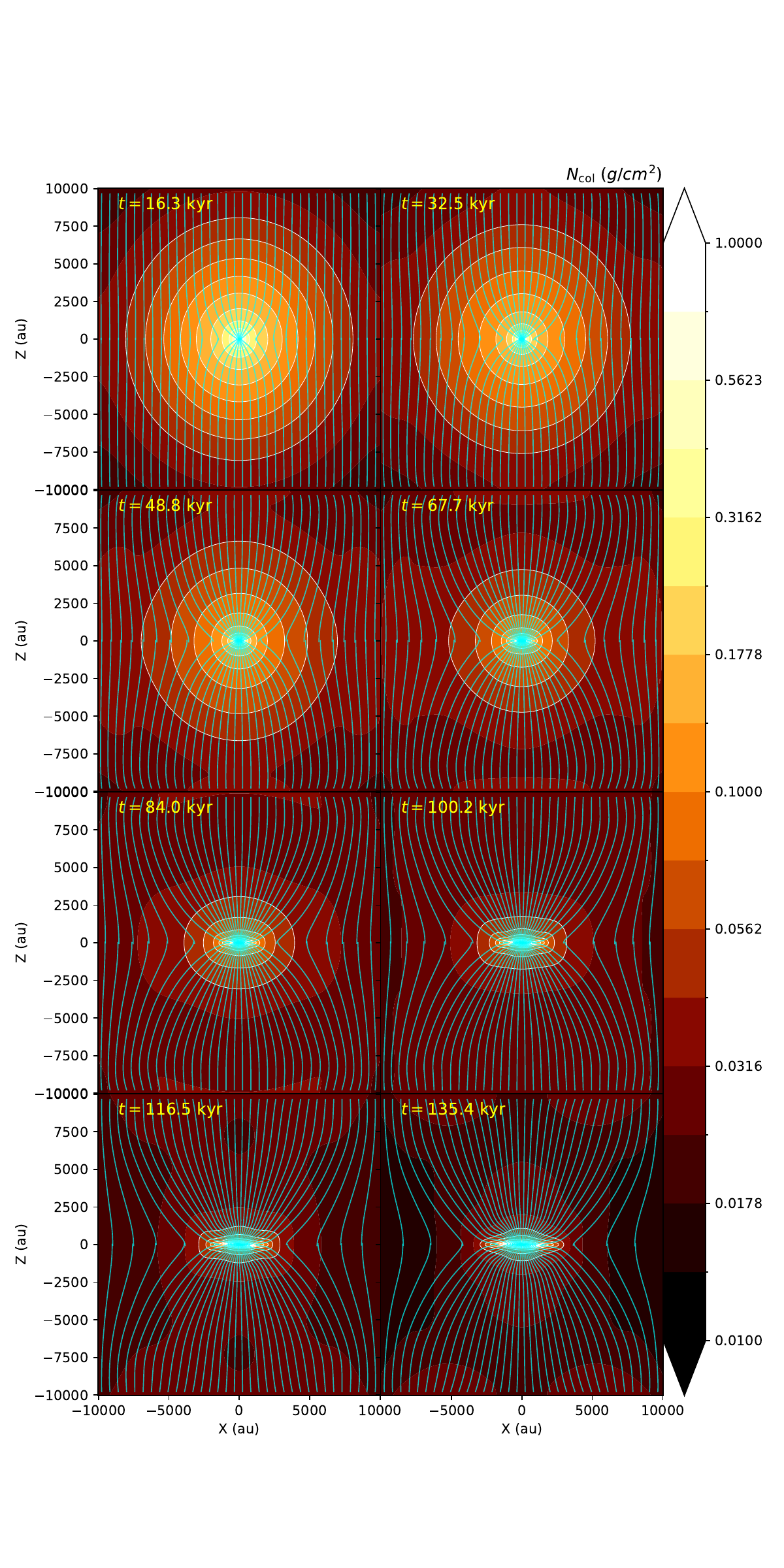}
\end{center}
\caption{Formation of a pseudodisk in time, $B_0 = 30 \muG$. The color and
white contours show column densities and the cyan lines trace the magnetic
field lines at the XZ plane. \label{fig:30muG}} 
\end{figure*}

\begin{figure*}[htb!]
\begin{center}
\includegraphics[width=0.8\textwidth, trim={0 3.3cm 0 4cm},clip]{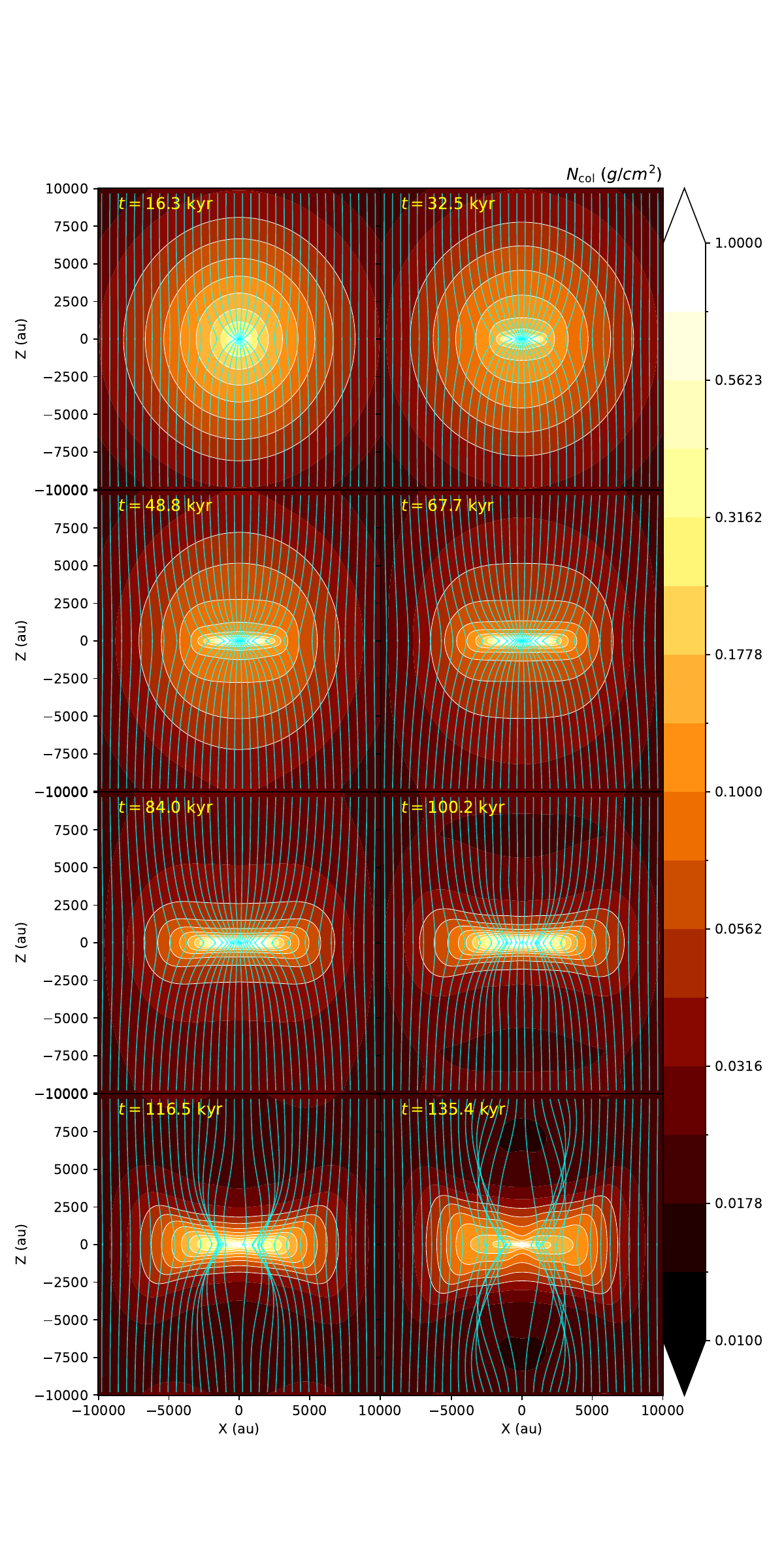}
\end{center}
\caption{Same as Figure \ref{fig:30muG} but with $B_0 = 150 \muG$. \label{fig:150muG}}
\end{figure*}

\begin{figure*}[htb!]
\begin{center}
\includegraphics[width=0.8\textwidth, trim={0 3.3cm 0 4cm},clip]{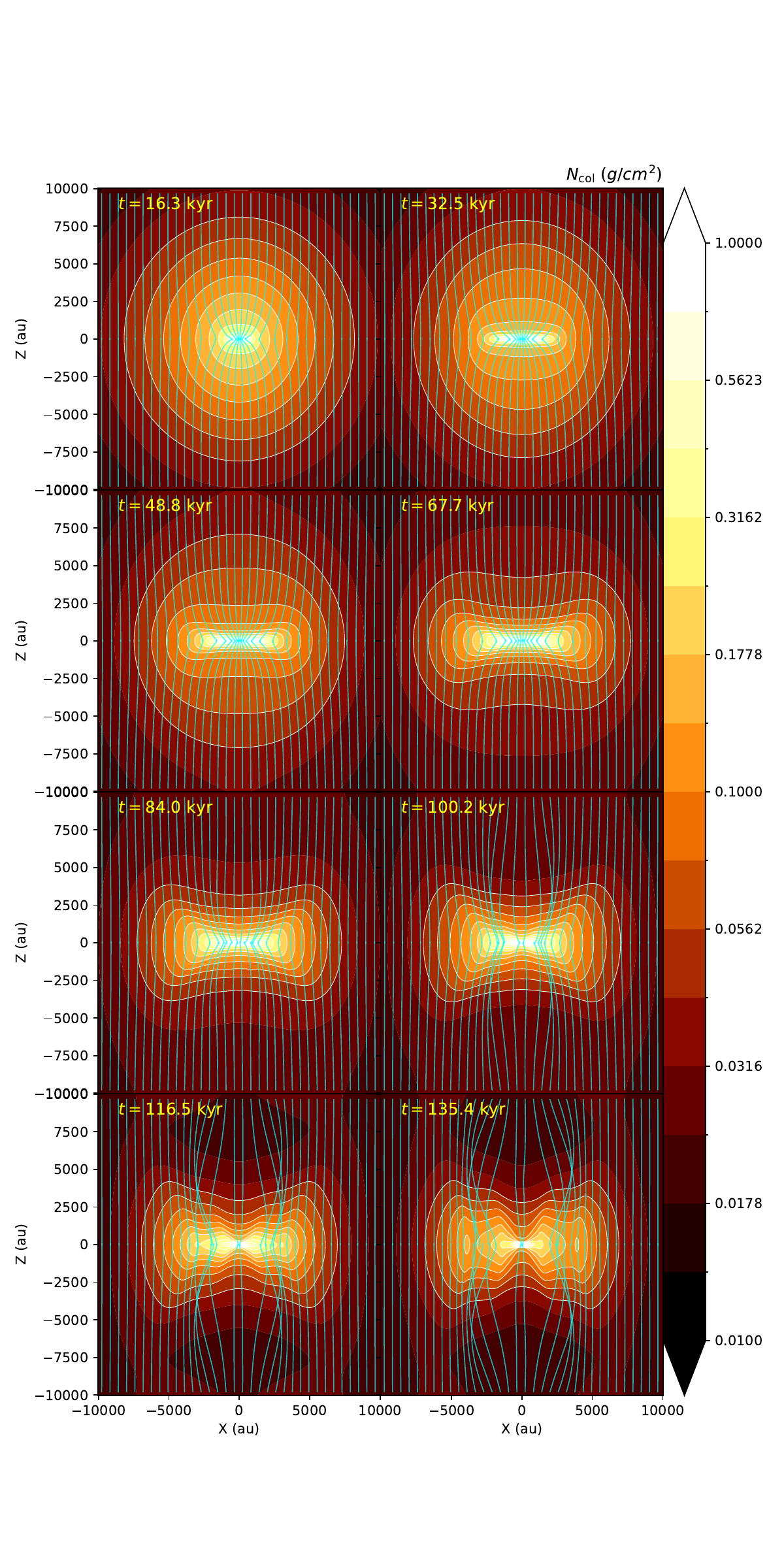}
\end{center}
\caption{Same as Figure \ref{fig:30muG} but with $B_0 = 270 \muG$. \label{fig:270muG}}
\end{figure*}

The panels in Figure \ref{fig:types} show the effect of increased magnetic
field on the shape of the pseudodisk formed at the stable end of the
simulations. Eventually interchange instabilities arise close to the sink
particles at the center. While these effects can be expected to occur in such
environment \citep{Ruben2012} we focus on the system behavior before the
instabilities happen. In general, the field around the centre becomes unstable
earlier, the stronger the initial magnetic field is. Figure \ref{fig:types}
shows the state of the system before this happens. Figures \ref{fig:30muG},
\ref{fig:150muG} and \ref{fig:270muG} show the corresponding systems evolving
in time. During the collapse, the isodensity contours appear first as
spheroids, then develop a flattened structure near the center of the system
while the larger-scale density evolve through toroid-like distributions (i.e.,
the center is thinner than edges of the pseudodisk). Apart from the general
toroidal shape, specific features of this density distribution are that the
central hole of the toroid is at the edge of the sink particle and that the
density increases toward the center \citep[see e.g.][]{LiShu1996}. The
$30\,\muG$ magnetic field case displays the flattened structure throughout the
simulation time. For larger field strengths, the density distribution evolves
through an increasingly flattened central region. In the $30\,\muG$ case with
very flattened isodensity contours, the center appears thicker than the outer
edges  due to the size of the sink particle. The late-time evolution of the
$B_0=150$~$\mu$G and $B_0=270$~$\mu$G cases (Figures \ref{fig:150muG} and
\ref{fig:270muG}) shows some concentration of field lines at intermediate
radii. This can be explained by the diffusive coupling of the magnetic field
and the inflowing matter, as near the inner regions matter is more prone to
decoupling from the field as it falls into the sink particle. Therefore, it is
possible for the outer portions of the field to fall in faster than some of the
inner regions, creating at the boundary an effect of diffusion-enabled pile-up
of the magnetic field lines
\citep{LiMcKee1996,CiolekKonigl1998,KK2002,MellonLi2009}. 

The system is under constant inflow toward the central gravitational
attractor. The density distribution is therefore constantly changing, far from
an equilibrium. The flow is subsonic at large distances from the center.
Eventually, the infall toward the central regions becomes highly supersonic,
reaching Mach numbers higher than 4, until it passes through an accretion shock
at the pseudodisk surface, where it becomes again subsonic (see discussion in
section \ref{sec:shockres}). 

\begin{figure*}[htb!]
\plottwo{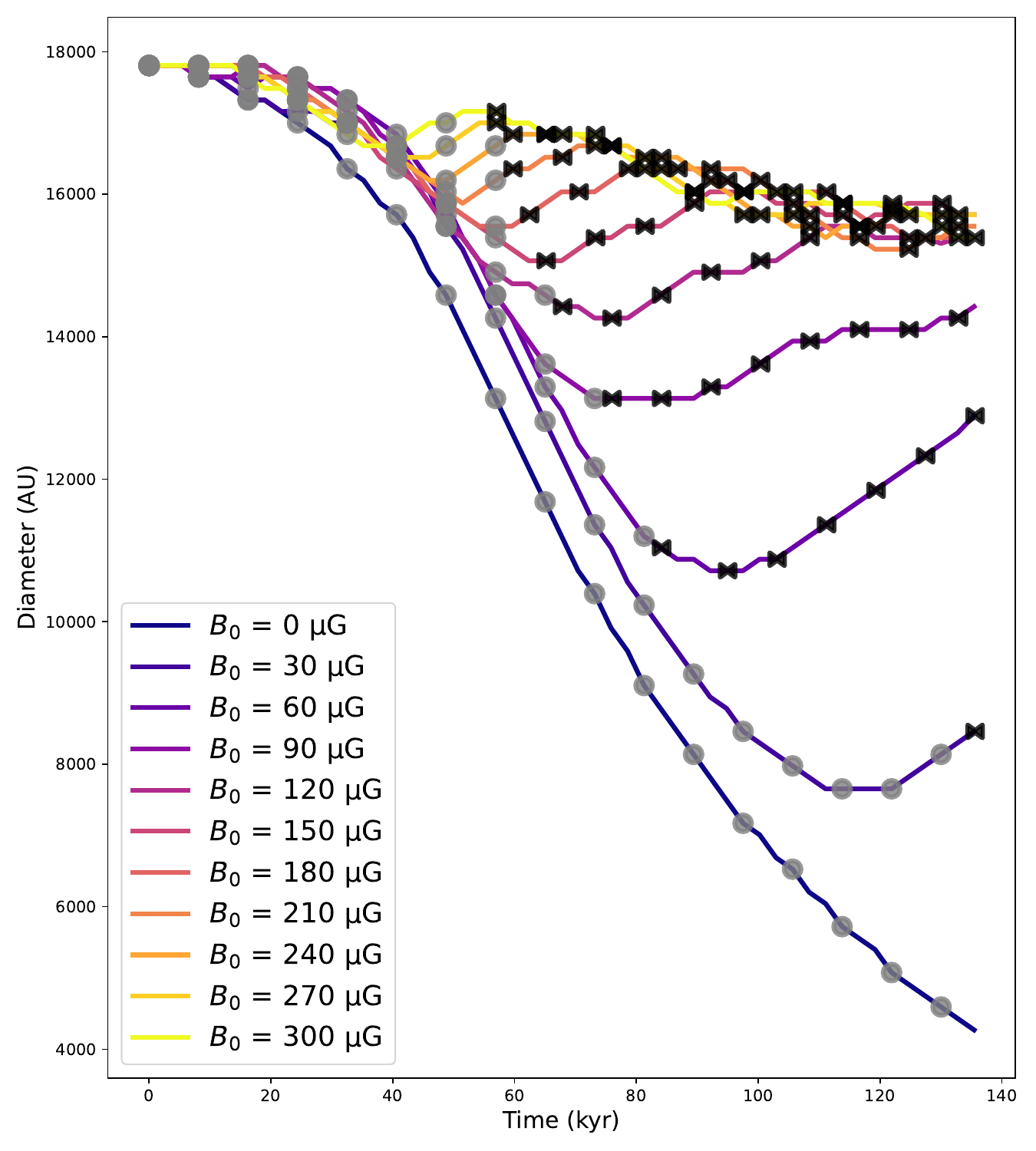}{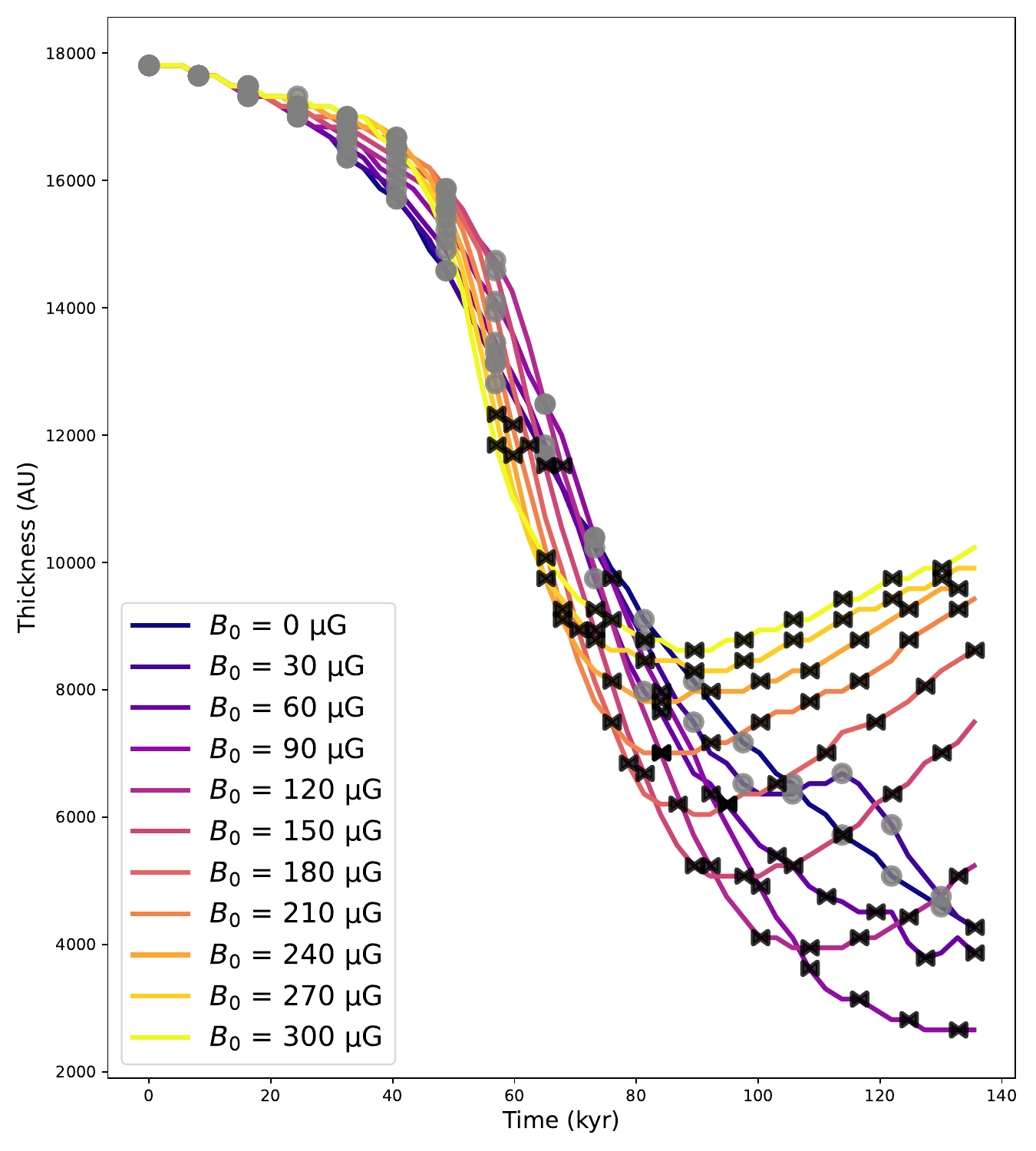}
\caption{The diameter (left) and thickness (right) of the collapsing core at
the threshold density $\rho_t=1.6\times 10^{-19}$~g~cm$^{-3}$. Filled gray
circles denote a spheroid shape, black crosses a toroid shape.}
\label{fig:ratio}
\end{figure*}

In order to quantify the flattening of the isodensity contours in time, we
measured the major and the minor axes of the contours at a selected threshold
density $\rho_t = 1.6\times 10^{-19} \gram~\mathrm{cm}^{-3}$. This threshold
density is arbitrarily chosen to highlight the shape of the flattened
pseudodisk Figure \ref{fig:ratio} displays the evolution of the pseudodisk
thickness and diameter in time, which are affected by the magnetic field
strength. Determining the pseudodisk diameter with a threshold density is
justified by observational criteria. In GS93II, the radius of the pseudodisk
was determined from the convergence of trajectories of fluid elements in the
ballistic approximation. In our simulations, however, the existence of the
isothermal pressure force term in equation (\ref{eq:momentum}), $-\css \nabla
\lnrho$, removes the sharp convergence of streamlines found by GS93II. 

Seen clearly in Figure \ref{fig:ratio}, the appearance of toroid features
corresponds to the stage where the diameter of the pseudodisk stops shrinking
and starts to increase. We note that the diameter of the pseudodisk starts to
increase earlier for stronger $B_0$. The minimum diameter corresponds to the
minimum of mass-to-flux ratio as shown in Figure~\ref{fig:masstoflux}. 
After this increase, and during the remaining evolution, the diameter has an
oscillatory behavior, reflected also in the mass, the flux, and mass-to-flux
ratio (see Figure \ref{fig:masstoflux}).

\subsection{Magnetic field lines} 

The behavior of magnetic fields during the collapse process is strongly
dependent on their initial strength, as coupling of the magnetic forces with
the flow governs the stiffness of the field. Strong magnetic force acts against
the collapse, and as it happens, the field is dragged to a lesser extent along
the flow. As expected, the magnetic field rapidly acquires an hourglass
morphology during collapse. The increase of the resistivity with density at the
center, softens the magnetic field kinks in this region.

\begin{figure*}[htb!]
\gridline{\fig{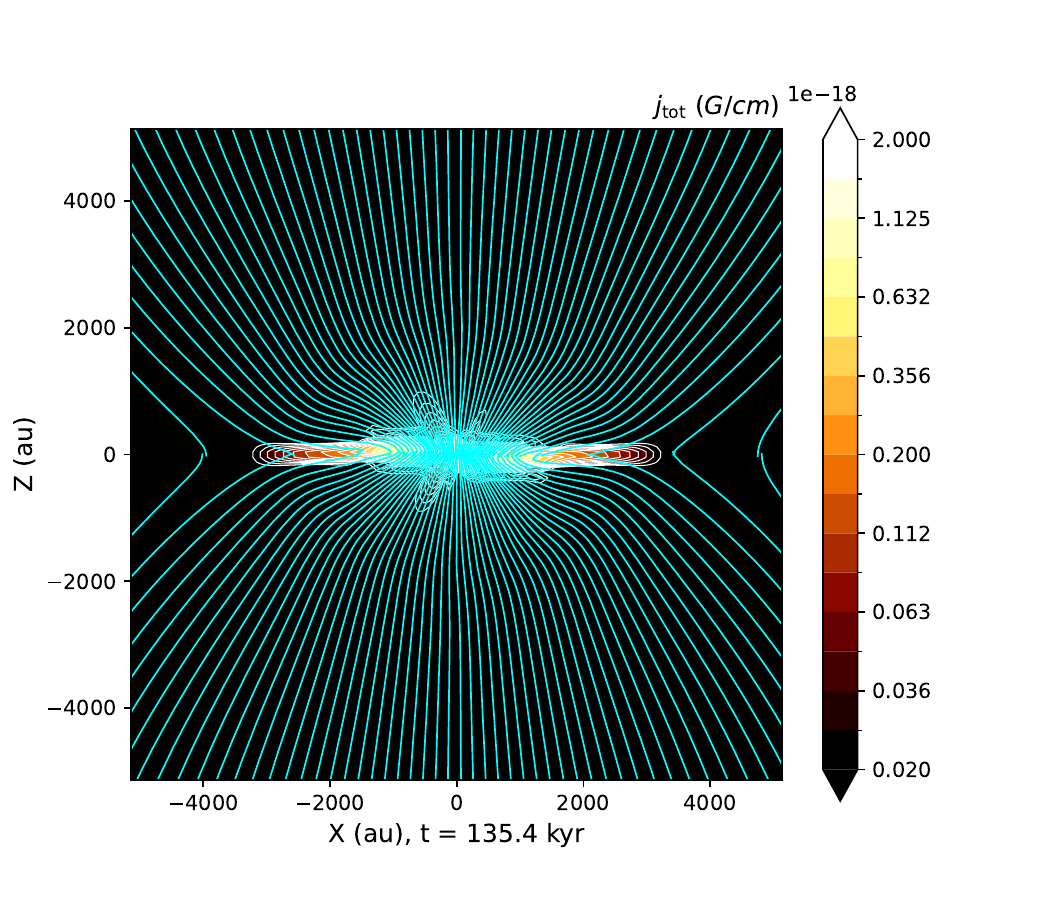}{0.5\textwidth}{(a) B030, $B_0 = 30 \muG$}
          \fig{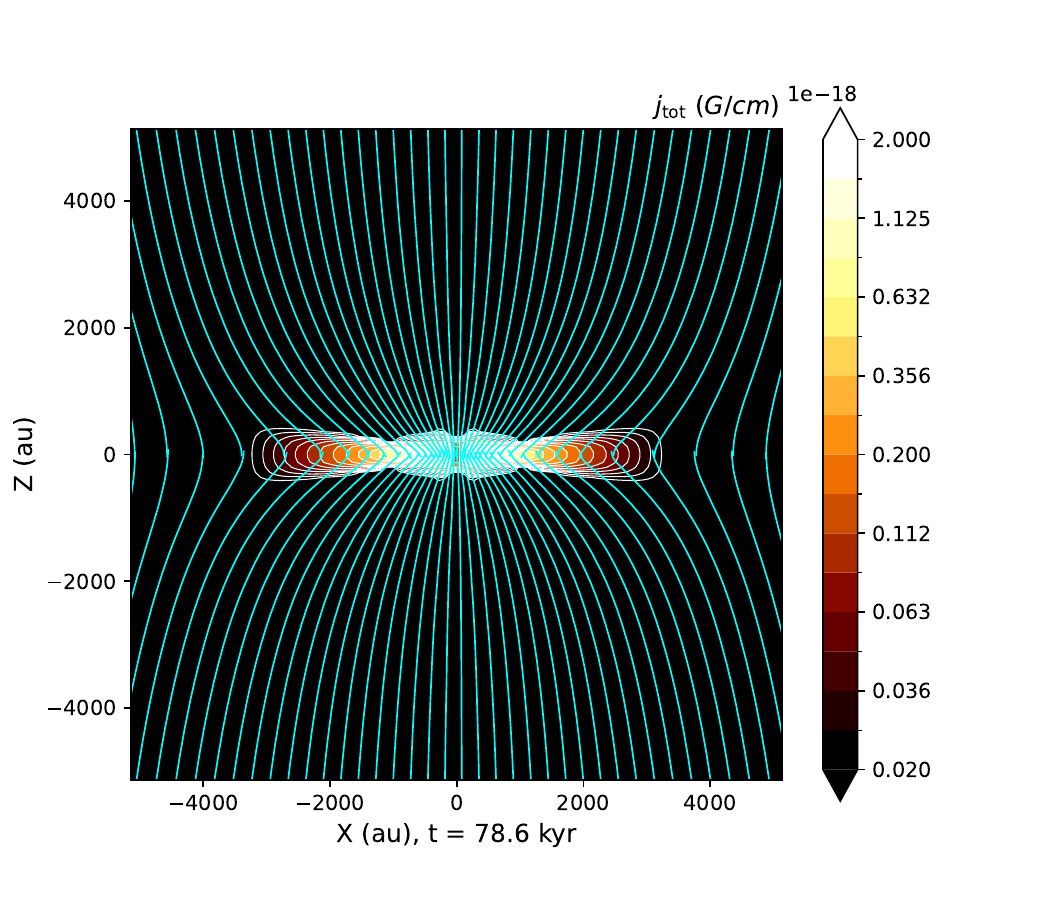}{0.5\textwidth}{(b) B150, $B_0 = 150 \muG$}
          }
\gridline{\fig{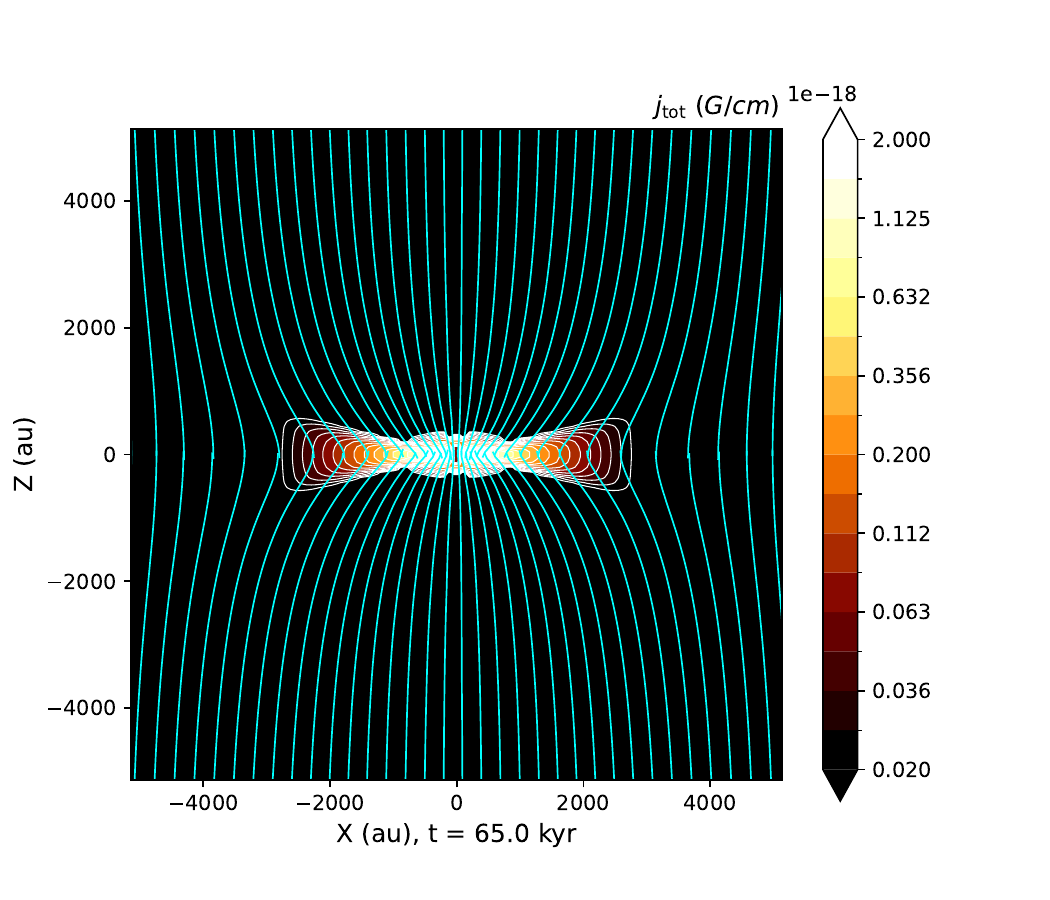}{0.5\textwidth}{(c) B270, $B_0 = 270 \muG$}
          }
\caption{As in Figure \ref{fig:types}, with cyan lines still tracing magnetic
field lines, but displaying electrical current density $j_\mathrm{tot} =
|\mathbf{j}|$ distributions, and the view is zoomed closer to the center. 
\label{fig:cursheet}
}
\end{figure*}

Given the presence of sharply pinched angles in the midplane, we inspect the
presence of current sheets. Calculating the electrical current density
$\mathbf{j}$, we find that $\mathbf{j}$ is generally distributed like a current
sheet as shown in Figure \ref{fig:cursheet} for the models shown in Figure
\ref{fig:types}. The current sheet becomes thicker and stronger with increasing
magnetic field, acquiring a bowtie shape
${\blacktriangleright\mspace{-10mu}\blacktriangleleft}$. 

The maximum magnetic field strength in the system reaches up to $\sim
12$--$14\,\mathrm{mG}$, with $B_0 = 30\mathrm{\muG}$. The systems with
initially weakest fields gather strongest fields over time as the field is
dragged toward the center. On the other hand, initially strong fields decrease
from their maximum value by several $\mathrm{mG}$ during their continued
development. 

\subsection{Mass Accretion Rates} 

\begin{figure*}[htb!]
\plotone{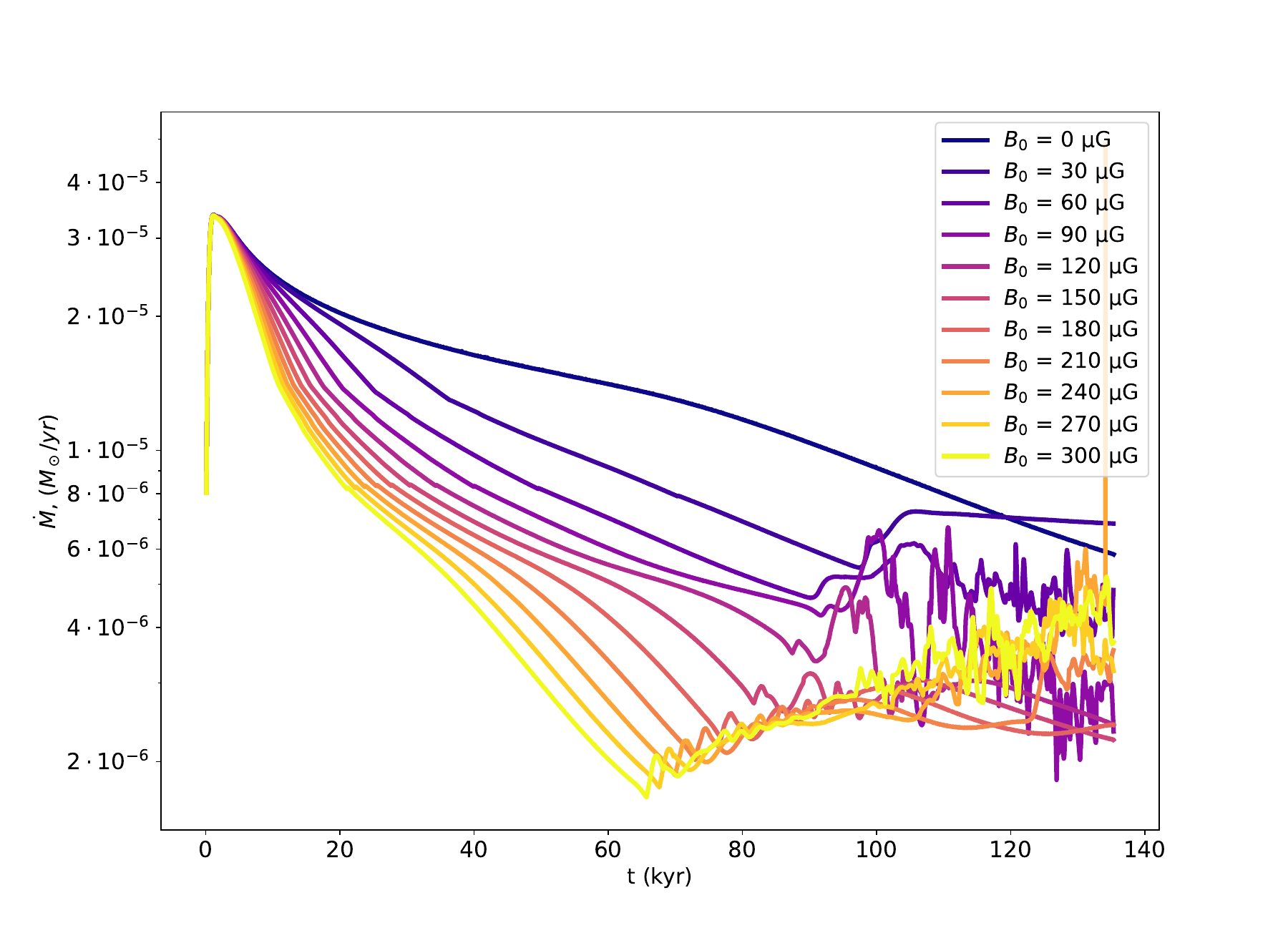}
\caption{Mass accretion rates  of all the systems as a function of time. 
\label{fig:massacc}}
\end{figure*}

Figure \ref{fig:massacc} shows the mass  accretion rate into the sink particle
as a function of time. In general, the accretion rate peaks soon after the
collapsing material starts falling toward the centre; then, it gradually
decreases. The strong initial accretion is an effect of the initial high mass
of the central point that drives the collapse. The decrease of the mass
accretion rate is also expected given the initial density profile that
decreases with radius, and that there is no mass inflow from the boundaries, so
the material runs out. Because our simulations do not start from equilibrium
condition, we can reasonably expect higher accretion rates than in
\citet{Shu1977}. The \citet{Shu1977} model predicts a constant accretion rate 
$\dot{M} = m_0 \mathrm{c}_\mathrm{s}^{3}/G = 0.975\mathrm{c}_\mathrm{s}^{3}/G$
\citep[See][Equation (8) and Table 1]{Shu1977} where in the case of our sound
speed, $\dot{M} = 9.9\times 10^{-6}\,\msun\,\yr^{-1}$. This is withing the
range of values of Figure \ref{fig:massacc}, but smaller than the most rapid
measured accretion rates. The accretion behavior is clearly affected by
magnetic field strength. Stronger fields decrease the accretion rate faster
after the initial peak because the stronger fields restrict the mass accretion
across the field lines. Toward the end, the accretion rate increases as the
magnetic systems reach unstable states, generally weakening the strength of the
field toward the central sink. However, in the case without magnetic field,
the gradual decrease of the accretion rate continues until the very end.

\subsection{Spherical Mass-to-flux ratio} \label{sec:masstoflux}

We calculate the nondimensional spherical mass-to-flux ratio $\lambda$ defined
as \begin{equation}
    \lambda = 2\pi G^{1/2} \frac{M(R_\Phi)}{\Phi(R_\Phi, 0)},
\end{equation}
following \citet{LiShu1996}. The flux $\Phi(R_\Phi, 0)$ was measured through a
circular surface at the equator ($z=0$) with a constant radius of $R_\Phi$. The
corresponding mass $M(R_\Phi)$ is the mass enclosed within a sphere of radius
$R_\Phi$, where $M(R_\Phi)$ includes both mass of the gas and of the central
object.

\begin{figure*}[htb!]
\gridline{\fig{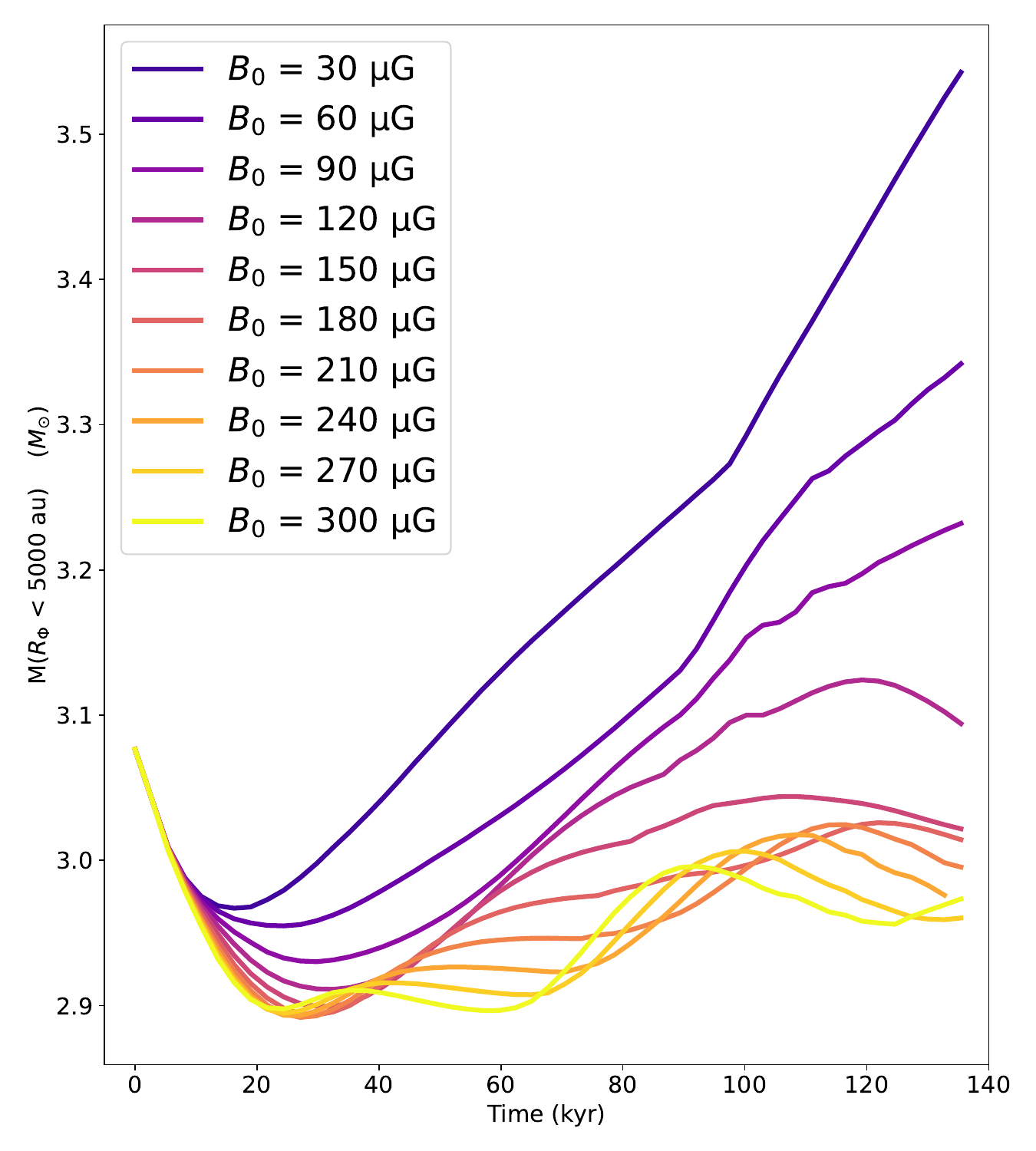}{0.5\textwidth}{(a)}
          \fig{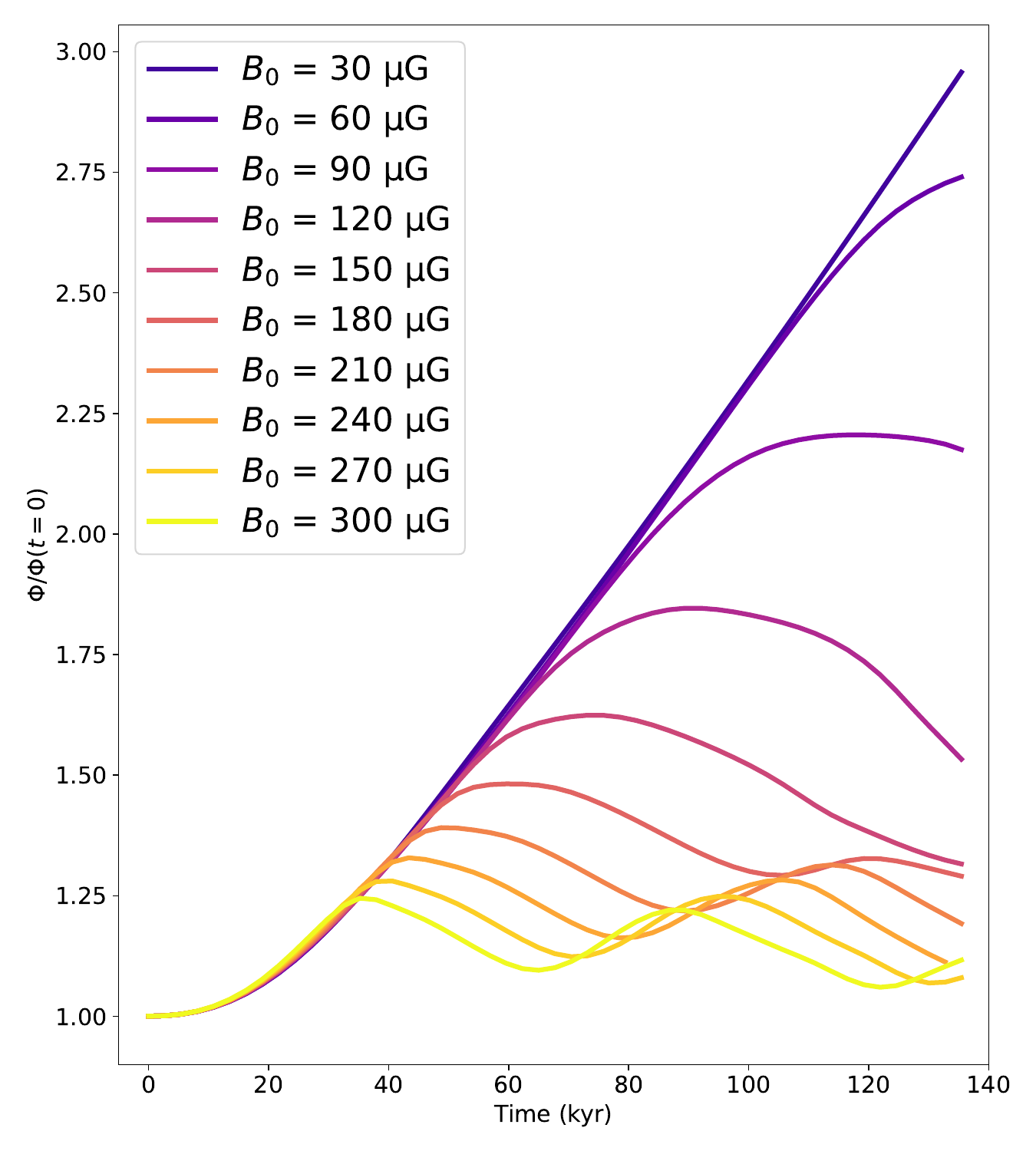}{0.5\textwidth}{(b)}
          }
\gridline{\fig{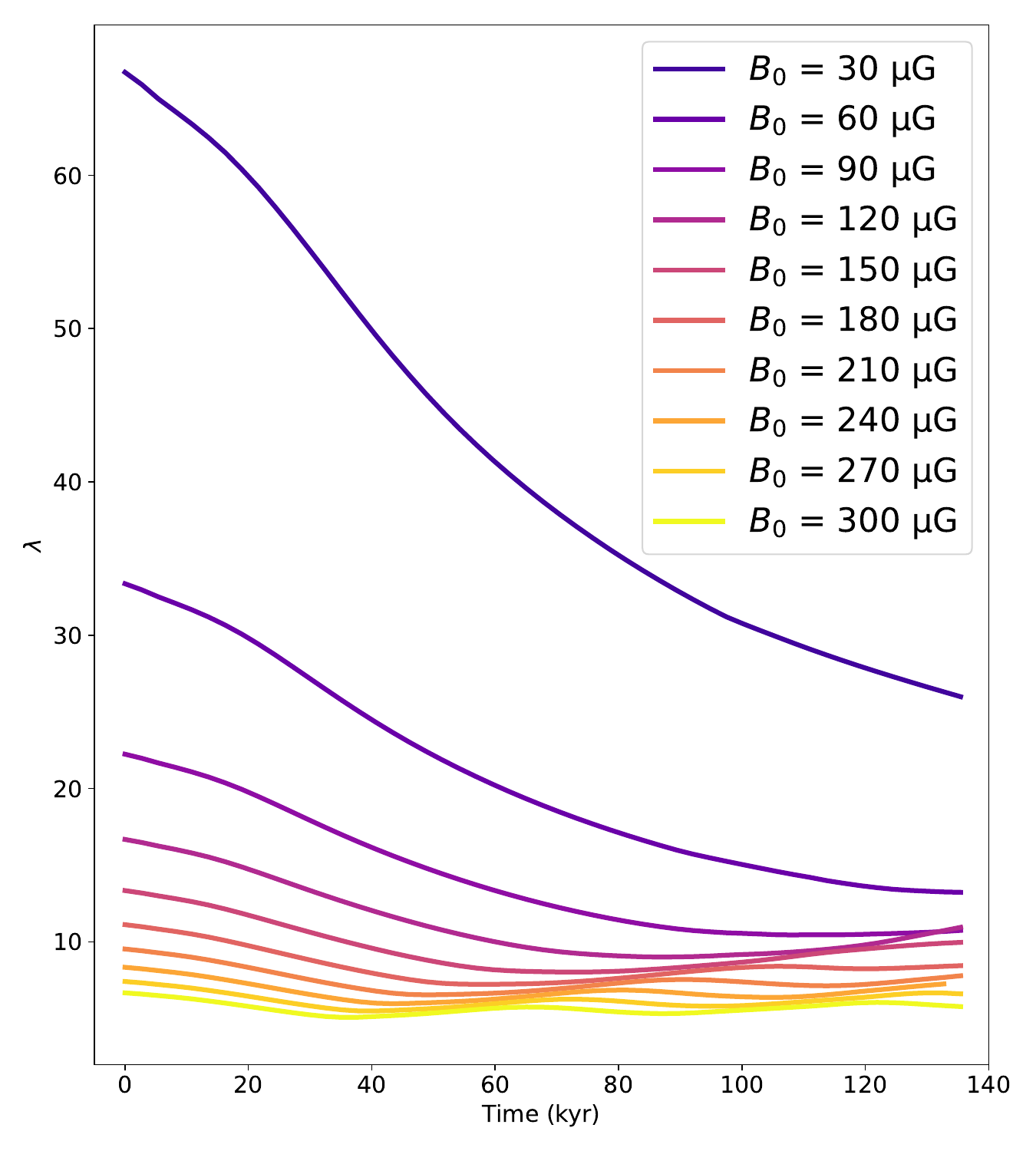}{0.5\textwidth}{(c)}
         \fig{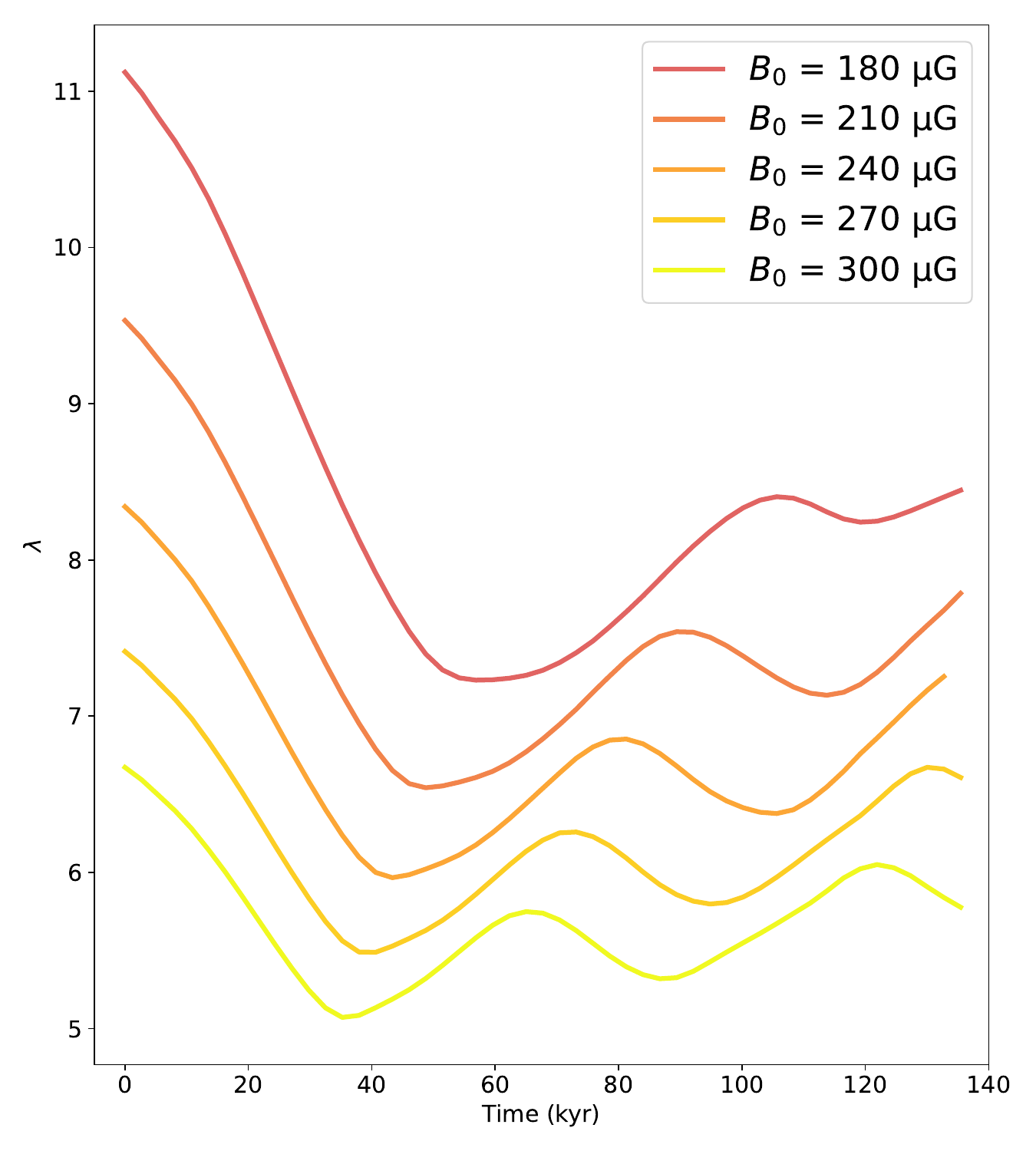}{0.5\textwidth}{(d)}
          }
\caption{Time evolution of the 
mass-to-flux ratio $\lambda$. The flux is calculated at the midplane within a
radius $R_\Phi = 5000\au$, for different values of the initial magnetic field.
Panel (a) a has the total mass within the $R_\Phi = 5000\au$ radius from the
center including the sink particle. Panel (b) shows  the evolution of the
magnetic flux with time. Panel (c) displays $\lambda$ for all $B_0$. Panel (d)
shows $\lambda$ for $B_0 = 180-300\,\muG$ to highlight the oscillation effect. 
\label{fig:masstoflux} }
\end{figure*}

Figure \ref{fig:masstoflux} shows the evolution of $\lambda$ for all systems
over time. We chose $R_\Phi= 5000\,\au$. Generally we see a linear decrease of
$\lambda$ from the initial state. This is followed by small amplitude
oscillations toward a more constant value. The decrease in $\lambda$ is due to
the assumed initial condition, where  the magnetic field is uniform and the
density is centrally concentrated, therefore $\lambda$ decreases with radius.
As accretion to the central source proceeds, it brings in material from the
outside with smaller values of $\lambda$. This is shown by the change in the
magnetic flux as a function of time (see Figure~\ref{fig:masstoflux}b). The
growth of magnetic flux is inversely influenced by the strength of the initial
magnetic field: the weaker the magnetic field, the stronger the flux can grow
relatively to its initial state. This is directly related to the fact that the
weakest magnetic field is the easiest to drag, with less magnetic forces
opposing the flow. In the case of $B_0 = 30\,\muG$ a maximum flux is not yet
reached within the simulation runtime. The oscillation of the flux and of
$\lambda$, particularly visible for the strong $B_0$, are caused by the back
reaction of the magnetic field on the gas, as the magnetic field resists to be
dragged to the center and the magnetic pressure builds up, causing the
temporary expansion of both the field lines and the gas. Thus, the dragging of
field and gas to the center happens in this oscillatory fashion (see
Figure~\ref{fig:masstoflux}d).

\subsection{Shock detection}
\label{sec:shockres}

As pseudodisks result from anisotropic supersonic inflow, it is relevant to ask
whether there are shocks present in the system. When examining our simulations,
we find a supersonic inflow that bends and collides close to the disk midplane
near the inner surfaces of the flattened density profiles. These regions would
be the places to look for shocks. Toward this goal, one can find the locations
of compression, or where there is a quick transition in velocity.

\begin{figure*}[htb!]
\begin{center}
\includegraphics[width=0.85\textwidth, trim={0 1.0cm 0 1.4cm},clip]{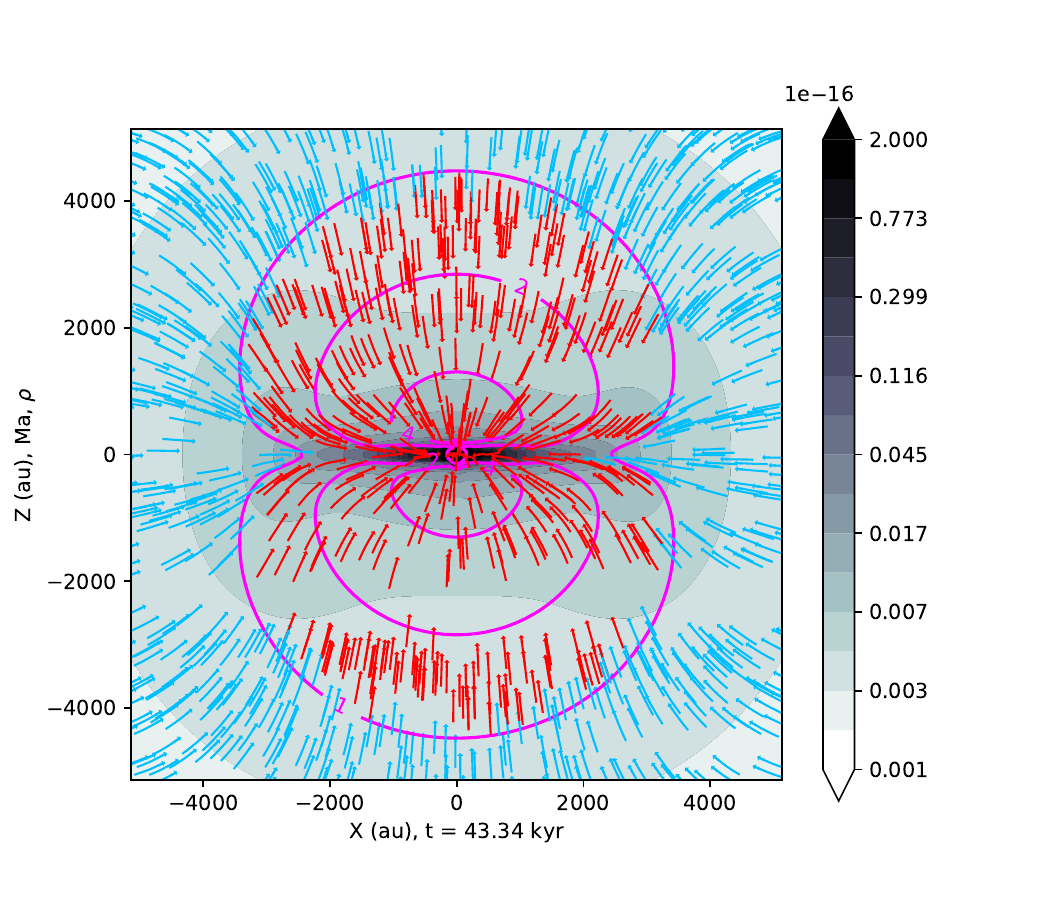}
\end{center}
\begin{center}
\includegraphics[width=0.85\textwidth, trim={0 1.0cm 0 1.8cm},clip]{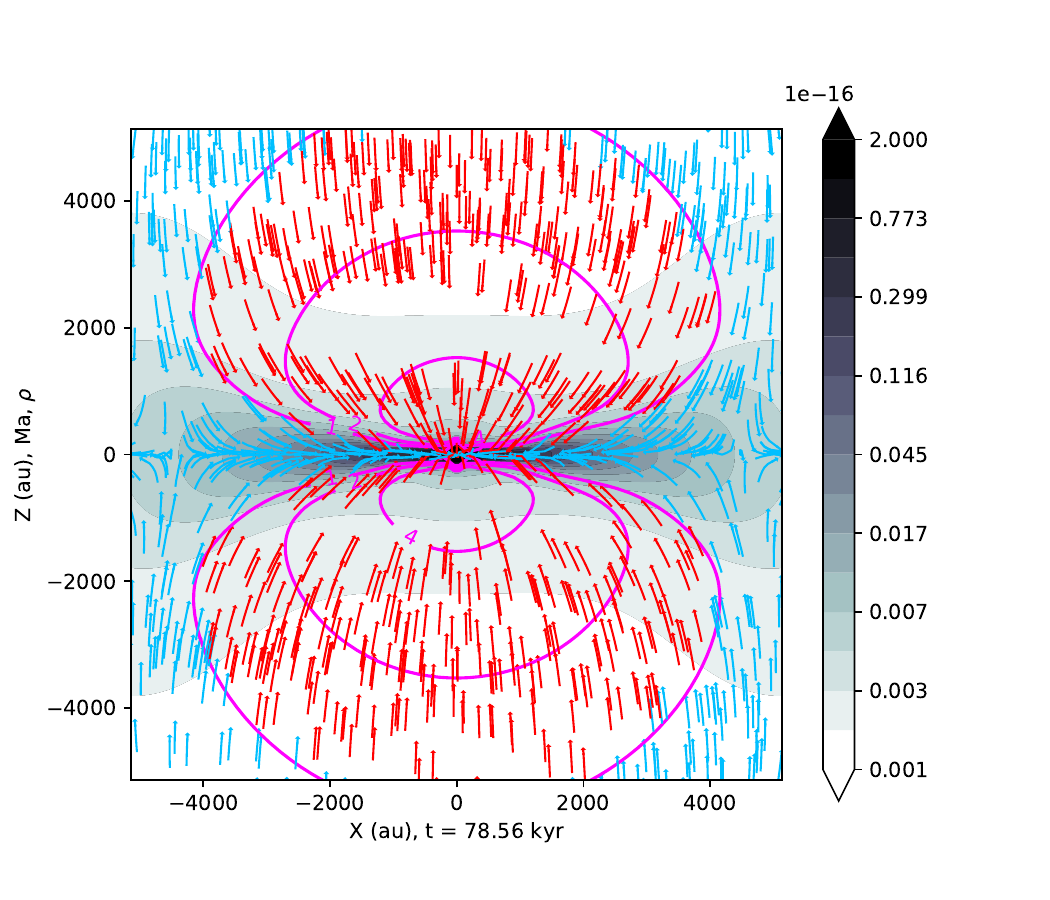}
\end{center}
\caption{Examining potential shock location in pseudodisk surfaces in B150 with
$B_0 = 150 \muG$ at early and late stages. The purple line contours show the
regions where the velocity at Mach numbers  1, 2, 4 and 8. The red arrows trace
supersonic flows, the blue arrows indicate subsonic speeds, and the grey
contours trace density. \label{fig:shocksample}}
\end{figure*}

\begin{figure*}[htb!]
\begin{center}
\includegraphics[width=0.85\textwidth, trim={0 1.0cm 0 1.4cm},clip]{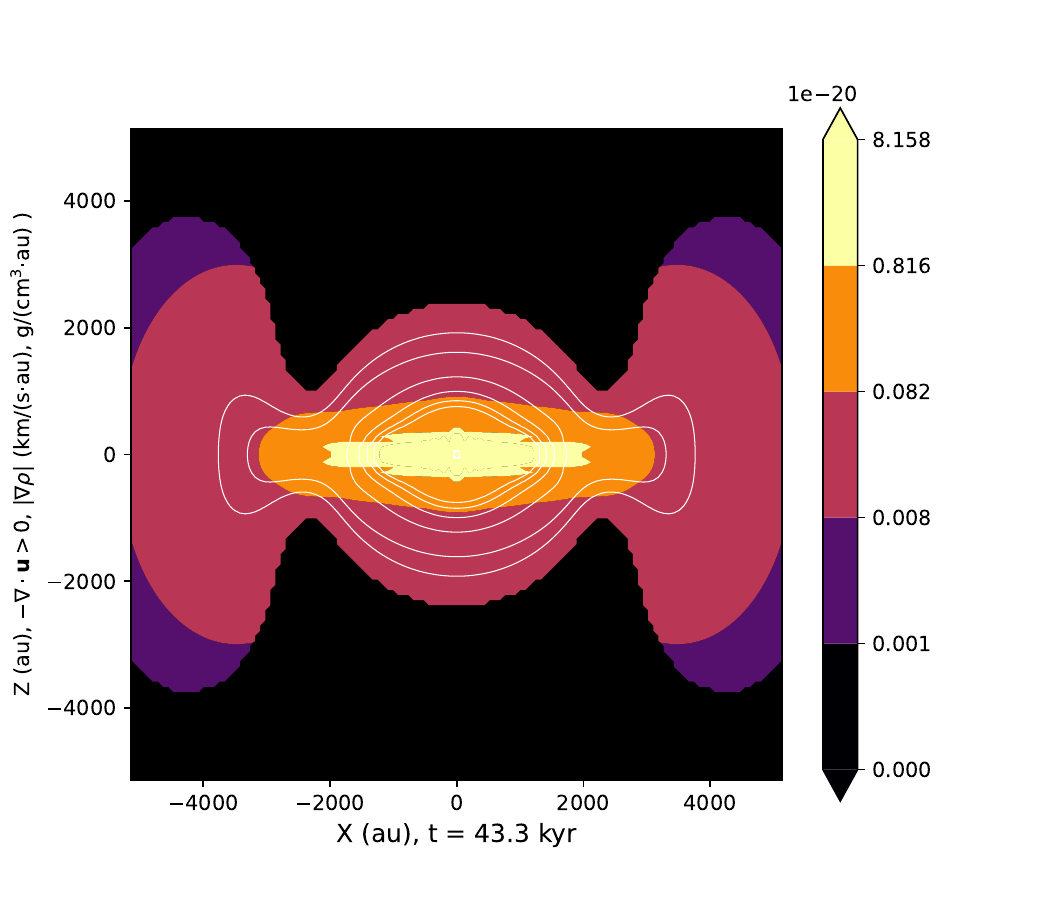}
\end{center}
\begin{center}
\includegraphics[width=0.85\textwidth, trim={0 1.0cm 0 1.8cm},clip]{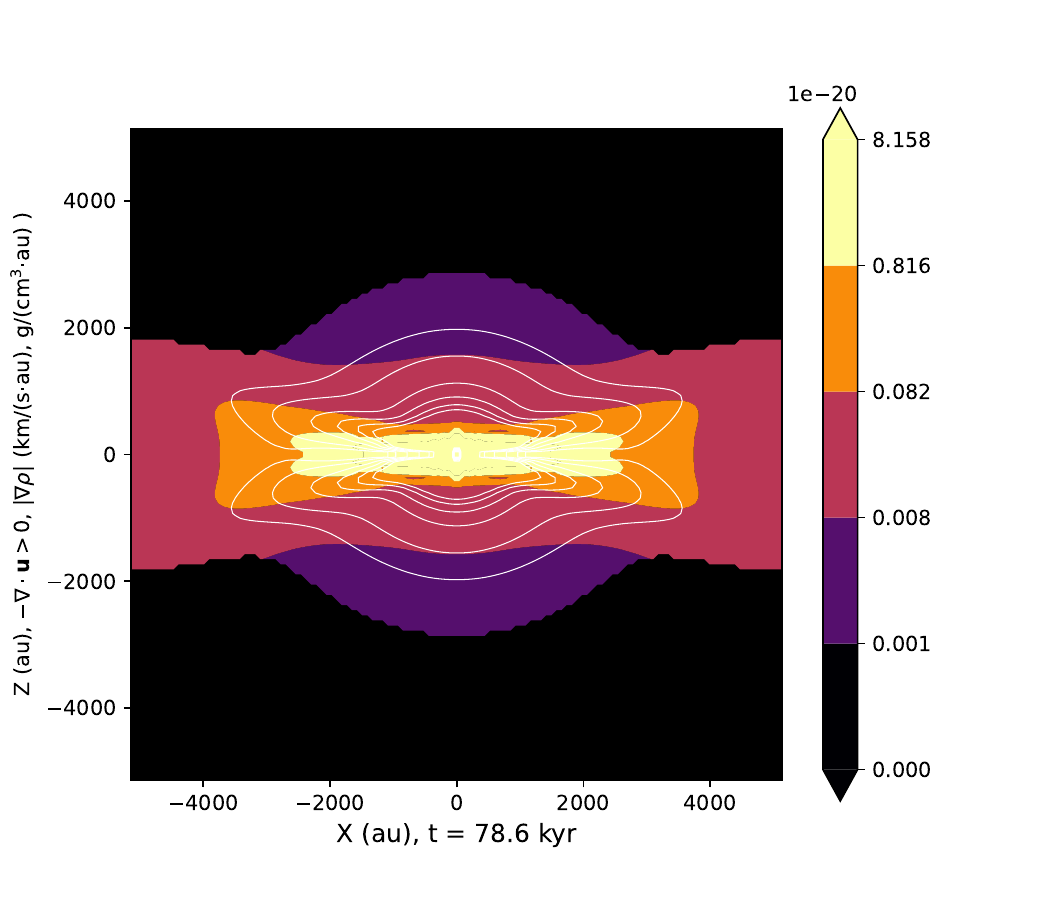}
\end{center}
\caption{Same frames as Figure \ref{fig:shocksample}, but displaying areas with
great velocity convergence (white contours) and density gradients (color
contours) as contributing tracer for inflow shocks. The white contours trace
the velocity convergence levels $\big[0.02 , 0.04, 0.09, 0.13, 0.17,
0.22\big]\,\km/(\second\,\au)$. 
}
\label{fig:shocksample2}
\end{figure*}

Figures \ref{fig:shocksample} and \ref{fig:shocksample2} show a representative
case of a shock detection. While details can be different, the general case
applies to all simulations where toroidal pseudodisks form. There is a general
supersonic inflow, which transitions from subsonic to supersonic at the edges
of expanding inside-out collapse wave. The flow converges toward the midplane
and the pseudodisk with a supersonic infall. Near the pseudodisk, shown in
Figure \ref{fig:shocksample} at 78.56 kyr, there appears a sharp transition
from supersonic to subsonic flow. This transition layer is a natural location
for shocks, and it appear as the toroidal pseudodisk forms. As shown in Figure
\ref{fig:shocksample2} these transition layers correlate with regions with a
high degree of compression as traced by both negative divergence of velocities
and density gradients.

Because these tracers -- deceleration of flow velocity from subsonic to
supersonic, density gradients and compression -- correlate on pseudodisk
surfaces, we judge that there are shocks present in our simulation. However,
the scale of these features is smaller than the grid scale, and cannot be
resolved.

\section{Code test and performance}
\label{sec:code_test}

\subsection{Comparison to ZeusTW model} 
\label{sec:zeustw}

To compare our results against a more established numerical scheme, we
constructed a 2-D axisymmetric model using ZeusTW \citep{Ruben2010, Ruben2012},
with properties comparable to those of our Astaroth model except in following
respects. First, the featured ZeusTW model is only two-dimensional and it uses
spherical coordinate system instead of Cartesian one. Therefore, it can be
expected to keep general symmetry better then the 3-D Cartesian simulation.
Second, it is a low order code, being more numerically diffusive. Third, the
central sink method is different as it is based on an inner boundary condition
for inflowing fluid. Therefore, the sink mechanism of Astaroth is within the
grid, whereas in  ZeusTW the sink is external to the grid itself.

\begin{figure*}[htb!]
\gridline{\includegraphics[width=0.5\textwidth, trim={0.0cm 0.9cm 1.8cm 0.0cm},clip]{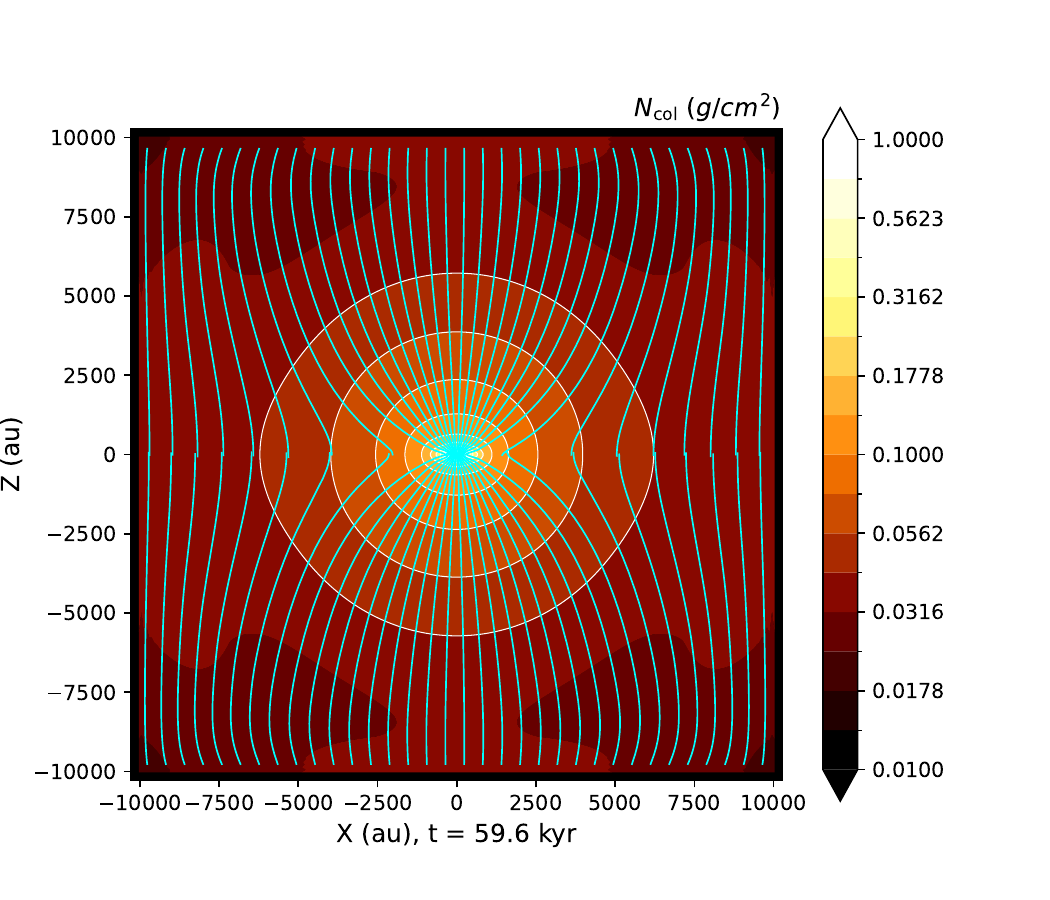}
          \includegraphics[width=0.5\textwidth, trim={1.1cm 0.4cm 0.6cm 0.4cm},clip]{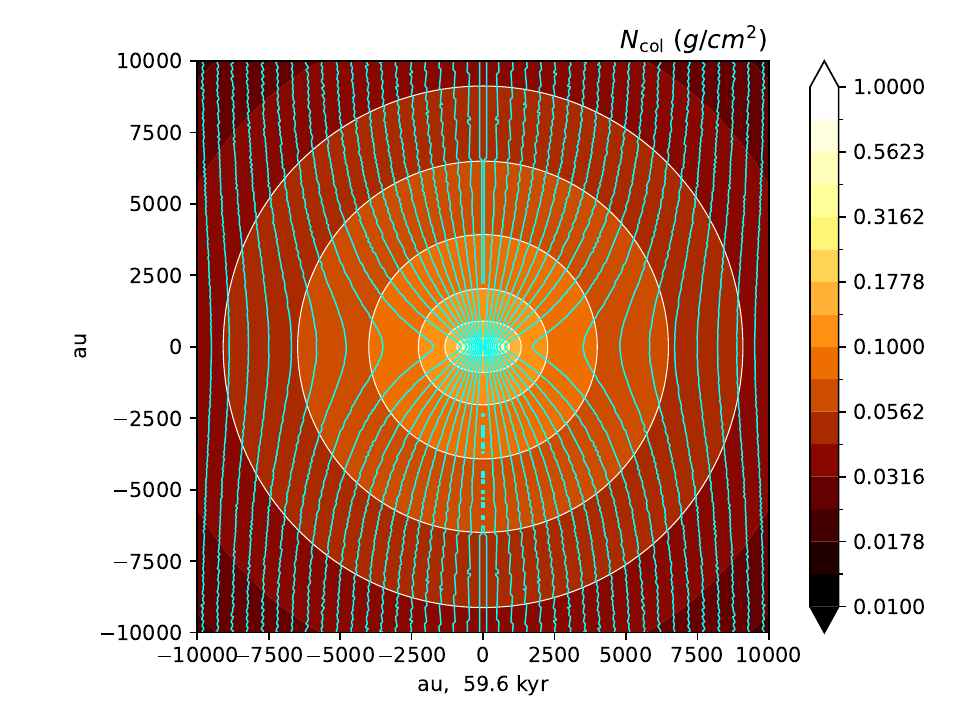}
          }
\gridline{\includegraphics[width=0.5\textwidth, trim={0.0cm 0.9cm 1.8cm 0.0cm},clip]{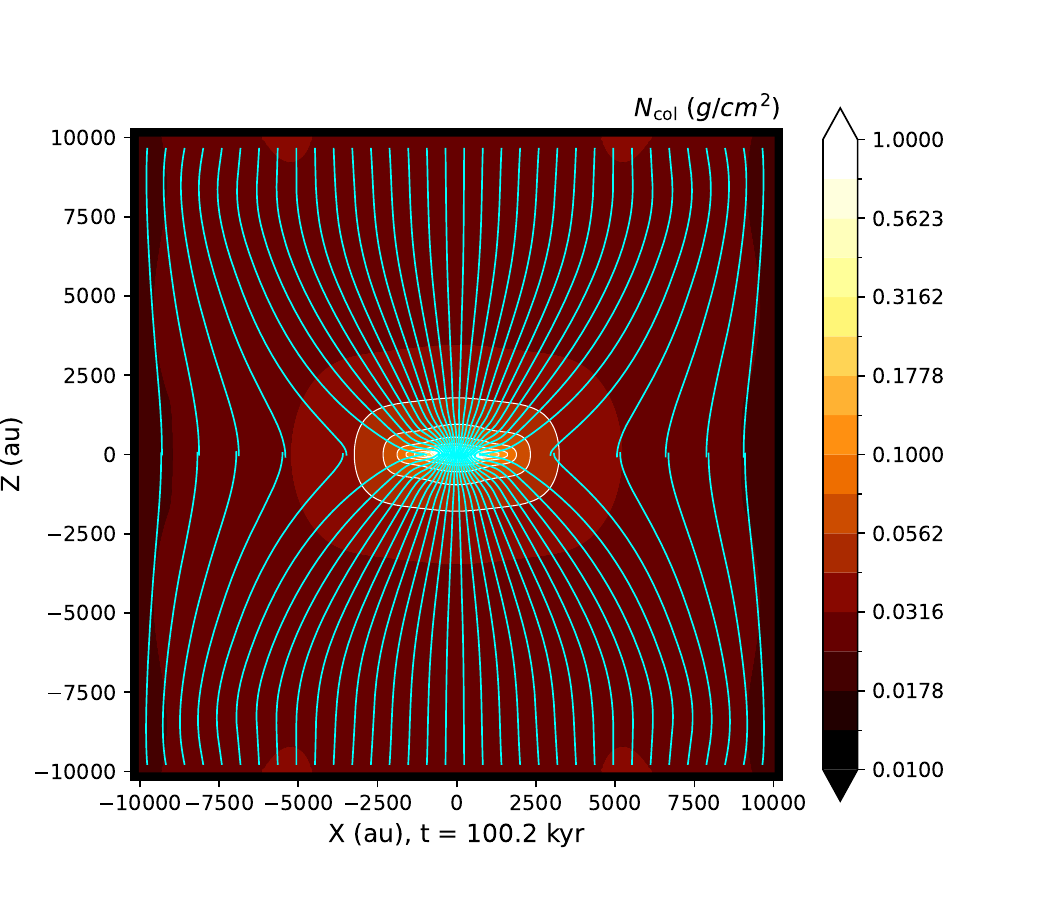}
          \includegraphics[width=0.5\textwidth, trim={1.1cm 0.4cm 0.6cm 0.4cm},clip]{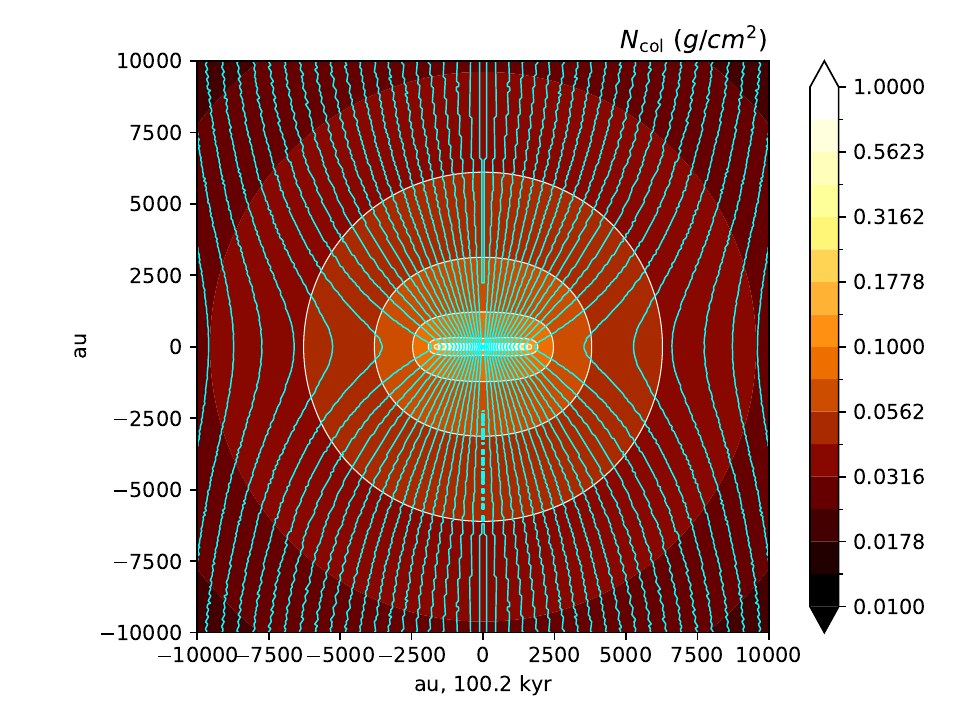}
          }
\caption{Left side shows Astaroth and right side ZeusTW+Perspective results at
corresponding times. Contours and curves represent column density and magnetic
field lines as in Figure \ref{fig:types}. There is difference in column density
contour levels because of different domain shape and size.
\label{fig:comparison} }
\end{figure*}

Figure \ref{fig:comparison} shows Astaroth and ZeusTW frames at a comparable
time. Column densities from the ZeusTW model have been calculated with
Perspective \citep{Vaisala2019}. There is a difference in the column density
levels, due to different domain size and shape, but local density levels are
comparable. The results are in qualitative agreement, and the time evolution
shows essentially the same progress, with pinching of the field and formation
of a thin flat pseudodisk. One difference is caused by the sink method, which
in Astaroth case creates a small bulge surrounding the sink particle.  Another
difference is in favour of Astaroth as the higher order code can handle a much
sharper pinch in the magnetic field, whereas the ZeusTW starts losing its
initial pinch.

\subsection{GPU Performance}
\label{sec:speedup}

Astaroth code performance has been benchmarked in \citet{Vaisala2021} and
\citet{pekkila2021}. They conclusively demonstrate the performance benefit of
the GPU implementation. For example, according to \citet{Vaisala2021}, a single
GPU device could perform about $10\times$ faster than a 40-core CPU node. Clear
performance benefit from the use of GPUs in comparison to more traditional CPU
parallel computing is to be expected, as it is has been demonstrated by most
GPU codes \citep[see e.g.][]{FARGO3D, Grete2021, Schive2018MNRAS}. With
Astaroth in particular, the high order of this computation adds to the stencil
size and, therefore, to the cost of each step due to memory access latencies;
which was challenging in the initial  development \citep{Astaroth2017} but has
been later solved \citep{Pekkila2019}.

\citet{pekkila2021} benchmarked scaling of the typical multi-GPU and multinode
Astaroth. When examining performance of Astaroth on a GPU \textit{node} they
measured that the speedup was 18--53$\times$ the performance of the Pencil Code
on a CPU node. When running the code on 16 nodes the gained speed up 
was 20--60$\times$ at best, using the largest grid size, with $N=1024^3$.
Other than speedup, they also compared energy efficiency with performance per
watt. In that case, as single node \textit{Astaroth} was 8--11$\times$ more
energy efficient then the Pencil Code CPU comparison. On 16 nodes, 
9--12$\times$ improved energy efficiency was measured at the largest grid size.
There reason why the largest problem size performs the best is because it
reduces the relative proportion of communication latency between multiple GPUs,
with system being closer to compute-bound than communication-bound conditions
when relatively large number of grip points are allocated per GPU.

However, those are not based on a completely identical setup to the one used
here, as our addition of shock viscosity and sink particle do add further
performance demands, due to shock viscosity in particular requiring multiple
stencil operation steps. Therefore, in this work we focus on examining the
code scaling with the properties we utilize. 

\begin{figure*}[htb!]
\plotone{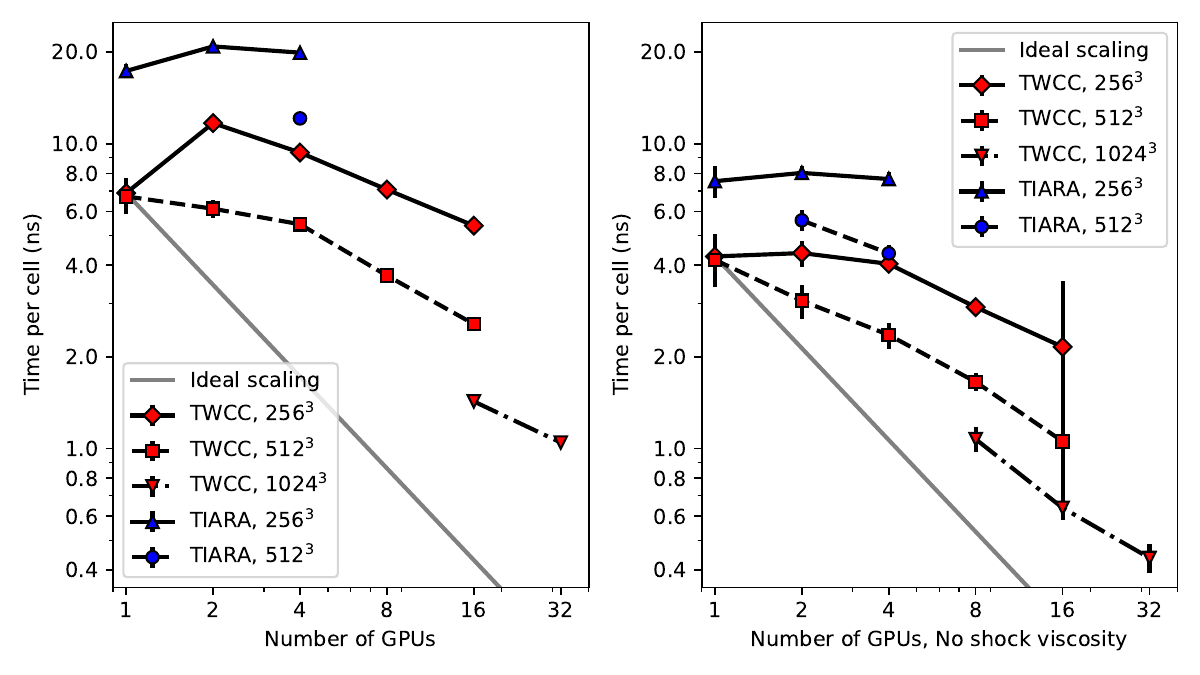}
\caption{Strong scaling. Performance is measured with fixed grid size while the
number of GPUs is changed, both with (left) and without (right) shock
viscosity. Error bars show the standard deviations of the measurement. Time per
cell refers to the time taken to calculate a single time step normalized by the
number of cells in the simulation. TWCC refers to Taiwan Computing Cloud, and
TIARA is the TIARA GP cluster of ASIAA. 
\label{fig:strongscaling}}
\end{figure*}

\begin{figure*}[htb!]
\plotone{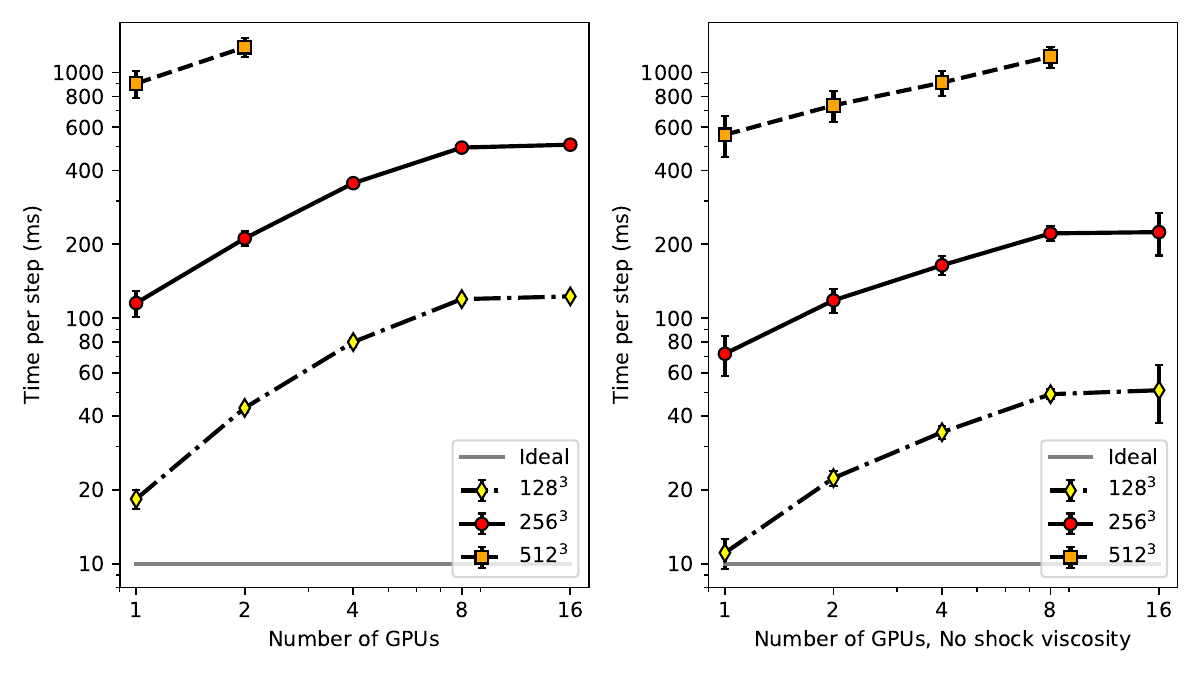}
\caption{Weak scaling, on TWCC. Problem size per GPU is kept fixed ($128^3$,
$258^3$ or $512^3$ per GPU), both with (left) and without (right) shock
viscosity. The estimated wall-clock time per step is shown without
normalization by the number of cells.
\label{fig:weakscaling}}
\end{figure*}

Figure \ref{fig:strongscaling} shows strong scaling with the computational loop
under a set of resolutions and available GPUs. First thing we can notice is
that with the primary resolution of the modeling work presented in this study,
$256^3$, a single GPU simulation run can be seen to be very economical. This is
due to the small grid and lack of device-to-device communication. While it is
true that at 16 GPUs the system is faster performance-wise, the gain is not
enough to justify the device count. However, many GPUs can be justified at
higher resolutions due to performance and memory requirements. At $512^3$ 8--16
GPUs could be considered a reasonable choice, and with $1024^3$ resolution, at
least 16 GPUs is essentially mandatory. The strong scaling follows comparable
slopes in all cases past 4 GPUs where doubling the GPU number provides about
1.4--1.6$\times$ increase in performance. In the most compute bound $1024^3$
case, without shock viscosity, scaling approaches about 1.7$\times$ speedup,
with the theoretical limit being 2$\times$ with the same increase of GPUs.

Benchmarks were performed with and without shock viscosity included in the
operations. Roughly speaking, the inclusion of shock viscosity would double the
needed computing time. This is likely because shock viscosity requires a series
of operations with several passes with device to device communication and very
large smoothing stencil. Regardless, the cost of the performance loss is worth
it due to much increased model stability. 

TWCC performs substantially better than the TIARA GP nodes. Reason for this
difference is in hardware, as TIARA GP has earlier generation P100 GPUs in use
while TWCC features more recent V100 GPUs with larger availability of parallel
GPUs overall.

Figure \ref{fig:weakscaling} shows the behavior of weak scaling. Weak scaling
follows the expected behavior where at first the added communication overhead
increases the required computing time, but this behavior starts to plateau at
higher number of GPUs. Weak scaling improves with larger problem size per GPU
in the range from $128^3$ through $\gtrsim 512^3$, which fit within device
memory.

Our numbers, however, are not exactly the same as the ones featured in
\citet{pekkila2021}. This is understandable, because we do not measure exactly
the same things. Our aim has been to process a full computational loop as it is
in the problem solved in this study. In addition to shock viscosity this
includes global reduction operations required for calculating the Courant time
step, and calculating the accreted mass to the sink particle.

Regardless, performance benefits can be experienced in a practical way. In our
case, one $256^3$ simulation took about ten hours on a single GPU to be
completed. With multiple available GPUs one could compute all our production
runs practically overnight. However, if we assume that $10\times$ speedup
estimate of \citet{Vaisala2021}, then similar operation would have taken more
than roughly 4 days on a series of CPU-nodes.

\section{Implications}\label{sec:disc}

\subsection{Observational considerations}
\label{sec:observation}

All of our models show a pinched magnetic field. To get a simple estimate of
how the field would be visible in a sub-millimeter polarization map, we
calculate the polarization of an edge-on core following the method of
\citet{Fiege2000}. With this method we calculate the column density $\Sigma$
and the corresponding Stokes parameters $Q$ and $U$, assuming an optically thin
system. $Q$ and $U$ are subsequently used to get a polarization angle $\chi$,
which is tilted 90 degrees to trace the projection of the plane of the sky
magnetic field. This type of calculation is only dependent on the integrated
column density and the magnetic field orientation. As such, the method of
\citet{Fiege2000} is a very simplified and should not be confused with more
sophisticated radiative transfer model. However, it can provide a quick
representation of a magnetically aligned dust polarization pattern. 

\begin{figure*}[htb!]
\plotone{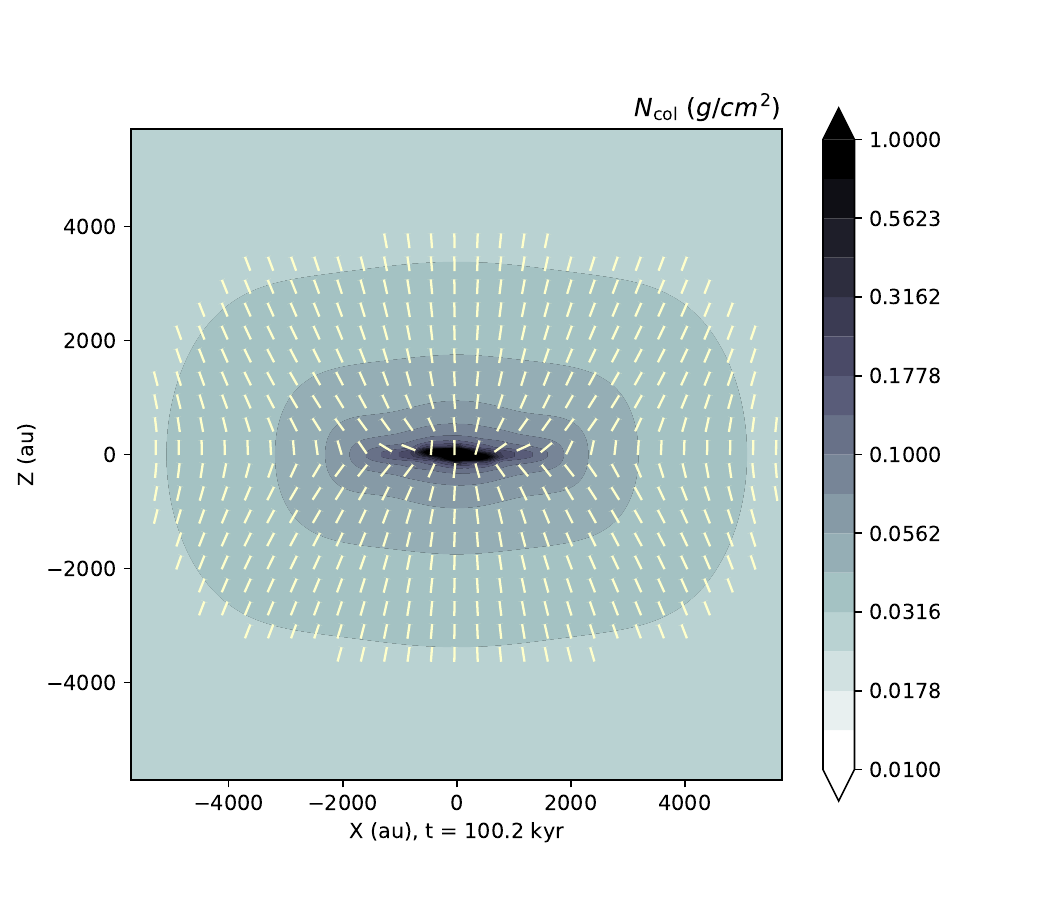}
\caption{Polarization map following \citet{Fiege2000} method, depicting the
case B030, with vectors matching the direction of the plane of sky magnetic
field. Polarization vectors are only shown over a column density threshold for
readability.
\label{fig:polpat}}
\end{figure*}

\begin{figure*}[htb!]
\begin{center}
\includegraphics[width=0.85\textwidth, trim={0 1.0cm 0 1.4cm},clip]{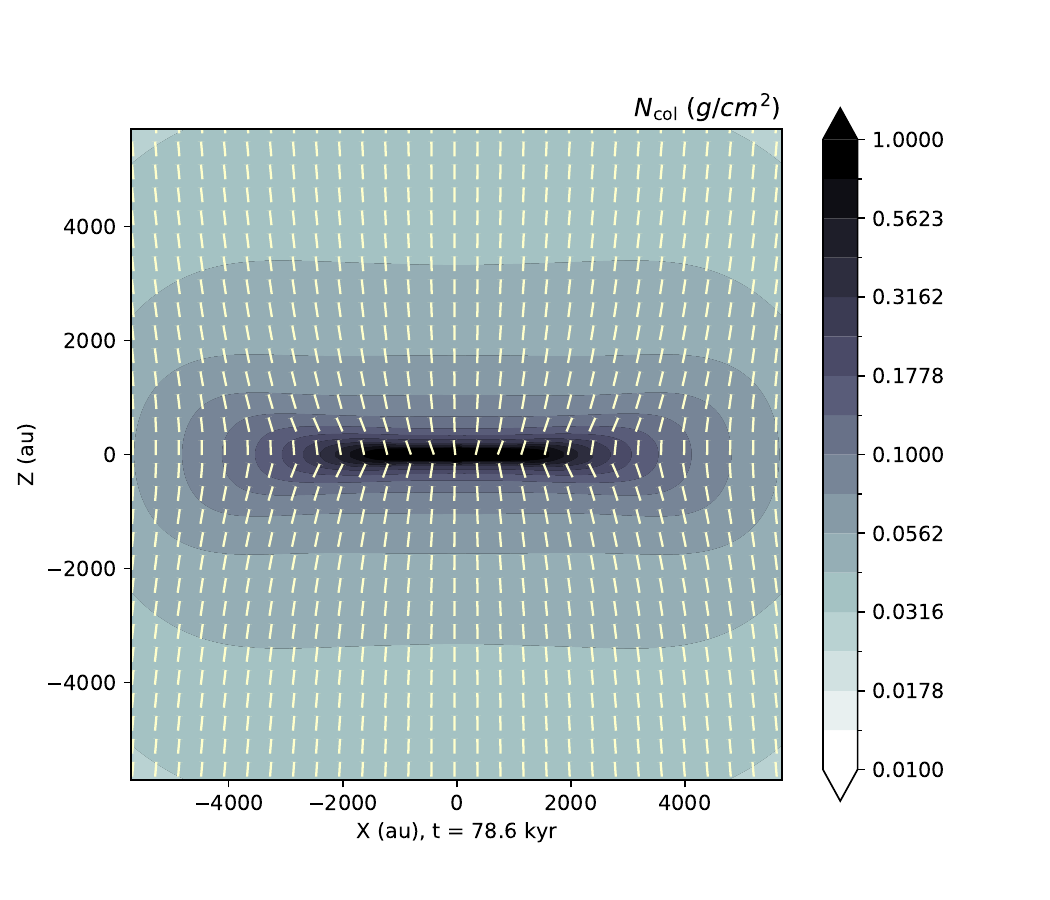}
\end{center}
\begin{center}
\includegraphics[width=0.85\textwidth, trim={0 1.0cm 0 1.4cm},clip]{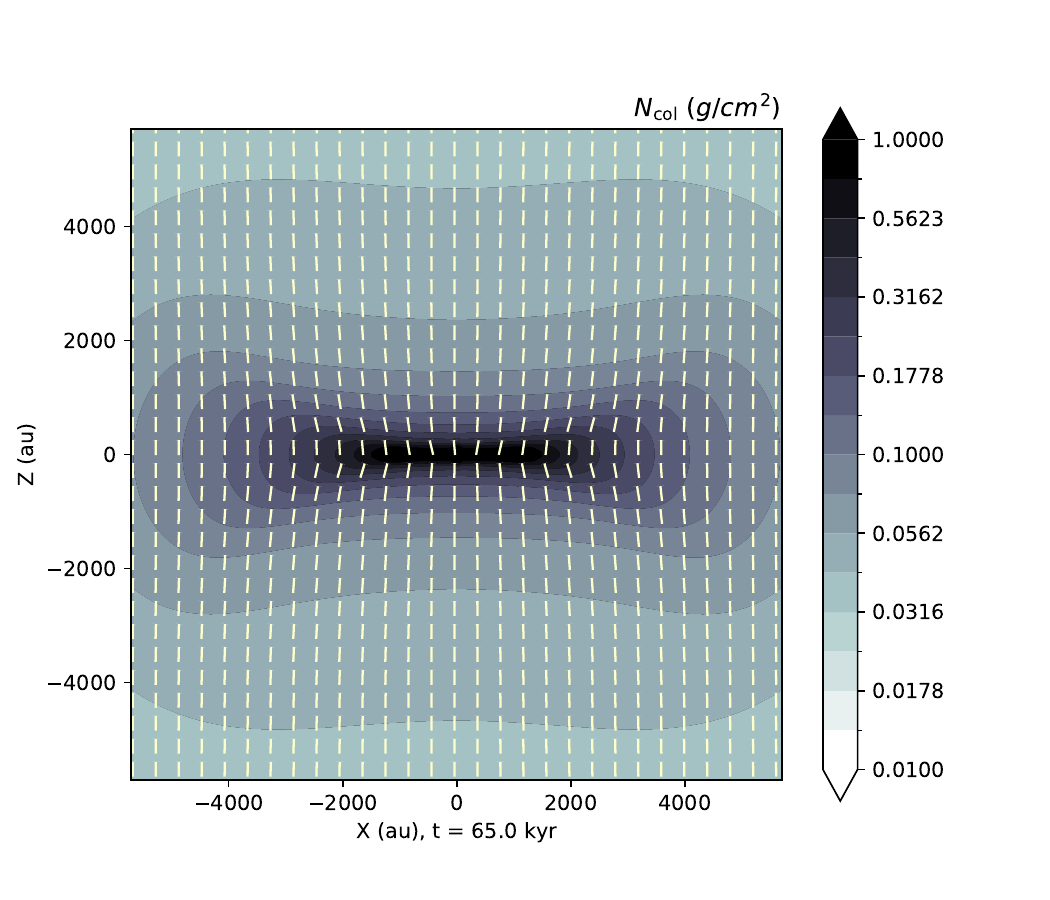}
\end{center}
\caption{Same as Figure \ref{fig:polpat} but with the cases B150 (top) and B270 (bottom).}\label{fig:polpat23}
\end{figure*}

Figure \ref{fig:polpat} features a typical example of a pinched hourglass
magnetic field seen in the polarization map calculated from our model.
Figure \ref{fig:polpat23} shows similar image for strong magnetic fields. The
pinch of the field become less visible for a larger value of the initial
magnetic field and the pinch is stronger toward the edge. Similar polarization
patterns are seen from various observations. 

Our model is focused on the pseudodisk and therefore its limits have to be
acknowledged. Our comparison can only apply to the protostellar envelope since
our model does not include rotation and, thus, it cannot produce a protostellar
disk or an outflow. 

Apart from HL Tau described in Introduction there are many other objects with
pseudodisk candidates, and are discussed below. Even though these observed
sources have protostellar disks and outflows, we are interested in discussing
their large scale pseudodisk structures.

NGC1333 IRAS 4A: a pseudodisk model has also been applied to the polarization
map of NGC1333 IRAS 4A \citep{GGG2008}. They concluded that the envelope of
NGC1333 IRAS 4A can be well fitted with a GS93I-type pseudodisk model from
which a magnetic field strength can be estimated. Recently, \citet{Ko2020} has
obtained further polarization maps at sub-millimeter frequencies. At large
scales, all the maps show the pinched magnetic field geometry, except for the
6.9 mm map where the polarization angle differs by 90 degrees. This difference
is explained as an effect of foreground extinction. 

L1448 IRS2: The magnetic field geometry derived from \citet{Kwon2019} has an
hourglass morphology and a substantially flattened density structure with size
scales matching pseudodisks, i.e., $\sim 1000 -2000$~au. Such large-scale
effects match well with a pseudodisk scenario, and being a Class 0 object,
L1448 IRS2 is at a stage where a pseudodisk would be expected to be present.
However, there have to be other contributions to explain the observed
depolarization close to the system's central plane. \citet{Kwon2019} state that
the depolarization can be caused e.g. by disturbances of the inner environment
by an outflow. However, both pseudodisk and outflow can exist in the same
system.  

L1157: This source has also the characteristics of a pseudodisk with an outflow
pushing out from the center.  \citet{Looney2007} discovered a flattened
envelope structure seen in infrared absorption surrounding the bright central
outflow. This structure is large and relatively diffuse in approximately 15000
to 30000 au in size. While this is larger than a pseudodisk of our model,
pseudodisk formation on larger scale is not uncalled for, as long as the
magnetic fields maintain a relative coherence with respect to the collapse.
Molecular line explorations of this object by \citet{Chiang2010}  discovered a
strong in-flowing velocity profile at the scale of the elongated envelope, as
is expected in pseudodisks. In addition, the inner envelope mapped at
$8\,\mu\mathrm{m}$ and in $\mathrm{N}_\mathrm{2} \mathrm{H}^\mathrm{+}$, being
close to size of $\sim 2000\,\au$, approaches the pseudodisk scales of our
models. This system potentially demonstrate the multiscale nature of the
pseudodisk phenomenon from more diffuse to dense gas. There is still accretion
toward the central protostar. The polarization observations by
\citep{Stephens2013} show an hourglass morphology of the magnetic field. 

B335: This source has polarization vectors that indicate a strongly pinched nad
organized hourglass magnetic field from from $1000\,\au$ to from $50\,\au$
scales  \citep{Maury2018}, which matches a typical pseudodisk scale. 
The radial velocity observations imply that rotation is weak at all scales,
with the majority of observed kinematics behaving as inflow \citep{Saito1999,
Yen2010, Yen2015}. \citet{Cabedo2022} found  that the ionization fraction is
sufficiently high to ensure a good coupling between the field and the gas.
Therefore, B335 appears as an almost  textbook case where the magnetic field
dominates the gas dynamics. This does not mean that there is no rotation, but
at large scales rotation is not the most dynamically relevant feature.

HH211: this source  displays the typical behavior of a magnetized collapse
\citep{Lee2019}: it has a collapsing inflow and a pinched hourglass magnetic
field. \citet{Yen2023} have performed further estimates and analysis on its
structure, kinematics, and magnetic field both at the envelope scale and at the
disk scale. The HH211 system  displays a pinched magnetic field configuration
with a flattened toroid-like envelope, with a significant inflow. Based on
their estimates, the magnetic field properties of HH211 are such that they
indicate a presence of a diffusive process such as ambipolar diffusion. 
The larger envelope still maintains a $r^{-2}$ density distribution, indicating
its relatively young age since the collapse has not produced sufficient
flattening on the largest scale. 

As can be seen from this list of candidate objects, it is possible to study
several pseudodisk-like protostellar envelopes. With further improvement, our
models will be able include both more complicated physics and a range of
physical scales, so that e.g. both a pseudodisk and a protostellar disk can
exist in the same model. This will allow us to address the details seen in the
observed systems more comprehensively.

\subsection{Implications to Early Models}

GS93I and II found the basic mechanism for the formation of a pseudodisk, where
the magnetic field that opposes the collapse and bends during the process,
deviates the gas flow toward the equator creating a disklike density
concentration. The GS93II simulation allowed for longer period of time,
compared to GS93I, where the pseudodisk flattened until artificial reconnection
effects near the center became too strong. However, it is noteworthy that while
GS93I assumes an isothermal equation of state, as does our model, the numerical
model of GS93II leaves the pressure force out of the calculations. Our
inclusion of isothermal pressure appears to lead to changes in the convergence
of inflow within the pseudodisk, as pressure maintains a finite scale height of
the density distribution. The effect is present even without any kind of
nonisothermal equation of state. Our low magnetic field cases are closest to
the GS93II results. In that case we get a relatively flat pseudodisk with
sharply bent magnetic field. With increasing magnetic field strength, the
pseudodisk is larger and thicker and the effects of pressure and compression
start to appear. Therefore, the density distribution has a more stratified
structure.  During the dynamical evolution of our models, there are stages in
which the isodensity contours are similar to the isothermal toroids of
\citet{LiShu1996} and to those found in the simulations of \citet{Allen2003I}
who studied the magnetized collapse of a cloud core without rotation, which
also lead to the formation of pseudodisks. Even though the mass-to-flux ratio
is different in these two latter works and in our simulation, it seems that the
toroid stage is a robust structure that appears during the evolution magnetized
cloud cores. 

\subsection{Infall Shocks}
\label{sec:shockdiscussion}

The existence for large-scale infall and accretion shocks during the
star-forming process has been a long-standing question.
\citet{YorkeBodenheimer1993, YorkeBodenheimer1995}, and \citet{
YorkeBodenheimer1999} presented the earliest cases in their hydrodynamic
simulations. These shocks are expected to produce  sharp density and velocity
contrasts. Recent observations have found shock signatures of infalling
material through the emission of SO and SO$_\mathrm{2}$ \citep{Garufi2022}.

In Section \ref{sec:shockres} we show the likely presence of infall shocks
in our systems with $B_0 = 150\,\muG$. The inflowing gas is strongly oriented
along the magnetic field lines: the low-density inflowing medium accelerates
along the field direction and becomes supersonic until it is slowed down and
deflected at the shock surface, where there are signs of both velocity
deviation and density compression. Detecting shocks within a supersonic system
where flows collide is not surprising. What is uncertain is the real strength
of the shocks, and how much thermal emission would be generated due to viscous
heating in those shock interactions. For example, \citet{ YorkeBodenheimer1999}
calculated the thermal emission of shocks formed by material inflowing onto a
disk surface. Such a calculation is beyond the scope of this paper.

\section{Summary}
\label{sec:conclusions}

We have performed simulations of the collapse of a nonrotating, uniformly
magnetized cloud core, using the high-order GPU code Astaroth. A pseudodisk
naturally forms for different strengths of the poloidal magnetic field. 
The collapse was examined with ten different levels of initial magnetic field
strength $B_0$, keeping the same initial density profile and central mass. Our
conclusions are summarized below.

(1) We find that density distribution evolves from spheroidal to toroidal
through the time evolution. The degree of central flattening of the final
configurations decreases with the value of the initial magnetic field. 
Toroids are favored in the case of stronger magnetic field, while
spheroids/flattened pseudodisks result in the case of weaker fields.

(2) The mass accretion rate on the central object increases rapidly at first,
and then decreases over time following a power-law type behavior, as the
surrounding cloud is depleted. At late times, mass accretion occurs in an
oscillatory fashion due to the back reaction of the compressed magnetic field
on the flow. The stronger the field, the faster is the decrease of the
accretion rate, and the earlier the oscillatory behavior sets in.

(3) The spherical mass-to-flux ratio $\lambda$ does not stay constant during
the evolution. Because of the initial nonuniform mass-to-flux ratio, the
buildup of the flux is much more rapid than that of the mass, leading to a
decrease in $\lambda$ with time. Furthermore, after the initial decrease,
$\lambda$ approaches a constant value with small oscillations, due to the back
reaction of the field on the gas. 

(4) Infall shocks are produced where the low density medium inflowing along
magnetic field lines accelerates to supersonic speed until it is slowed down
and deflected at a shock located at the surface of the pseudodisk.  

We find that the code Astaroth can now produce pseudodisk from a basic
principles, and this can function as a foundation for more sophisticated models
in the future, where additional physical processes can be added. For example, a
further implementation of a Poisson solver can allow for inclusion of
self-gravity. Additionally, more complicated nonideal MHD mechanisms like
ambipolar diffusion can be introduced.  Rotation can also be included to follow
the formation of a centrifugally supported disk at a a scale much smaller than
the pseudodisk. To investigate these inner protostellar disks, the resolution
has to be increased and the grid made nonequidistant. 

\begin{acknowledgments}

The authors would like to thank Johannes Pekkil\"a, Oskar Lappi, Matthias
Rheinhardt and Maarit K\"apyl\"a of Aalto University, Finland, for their
assistance and contribution for implementing the shock viscosity feature in
Astaroth, and their help with benchmarking. The authors appreciate Kouichi
Hirotani for useful discussions. 
The authors thank the careful reading and suggestions of an anonymous referee,
which improved the clarity and presentation of the paper.
The authors acknowledge the access to high-performance facilities (the TIARA
cluster and storage) in ASIAA, and thank the National Center for
High-performance Computing (NCHC) of National Applied Research Laboratories
(NARLabs) in Taiwan for providing computational and storage resources.  
M.S.V., H.S., and R.K. acknowledge grant support for the CHARMS group
from the Institute of Astronomy and Astrophysics, Academia Sinica
(ASIAA), and the National Science and Technology Council (NSTC) in Taiwan
through grants 110-2112-M-001-019- and 111-2112-M-001-074-. 
S.L. acknowledges support from grant PAPIIT/UNAM IN103921.
Astaroth is open-source under a GPL 3 license and it is available  at
\url{https://bitbucket.org/jpekkila/astaroth/}. This work utilized tools
(Astaroth, Perspective) developed and maintained by the CompAS and CHARMS
groups, and is part of the Astaroth collaboration network.
This research has benefited from the SAO/NASA Astrophysics Data System.

\end{acknowledgments}

\software{NumPy \citep{harris2020array},
SciPy \citep{SciPy-NMeth2020}, Matplotlib  \citep{Hunter2007MPL}, 
Pandas \citep{reback2021pandas,mckinney-proc-scipy-2010}}

\appendix

\section{Shock viscosity}\label{sec:shockvisc}

In addition to resistivity and kinematic viscosity, an artificial shock
capturing viscosity is needed. Shock viscosity is a required property near the
center of the collapse where velocities can get highly supersonic. However, it
has little effect on the general behavior of the collapse process, with the
effects of shock viscosity being local and not global. In Equation
(\ref{eq:momentum}), $\zeta_\mathrm{shock}$ is an artificial viscosity
parameter computed as follows. It is based on the method used by the Pencil
Code and presented by \citet{Gent2020}. First, we get a scalar field $u_{-D}$
by calculating negative divergences while setting every nonnegative diverge
value to zero. 
\begin{equation}
u_{-D} = (-\nabla \cdot \uu)_+
\end{equation}
Second, we get a scalar field $u_{5}$ by picking maximum within the points in
\begin{equation}
u_{5}(i,j,k) = \mathrm{max}[u_{-D}(i-2\ldots i+2,
j-2\ldots j+2,
k-2\ldots k+2)],  
\end{equation}
where $i$, $j$ and $k$ are the local grid indices.
As the last step we perform smoothing on $u_{5}$ with a normalized 
$7 \times 7 \times 7$ window with Gaussian weights 
$w(i) = [1, 9, 45, 70, 45, 9, 1]$ in each direction to get scalar field
$f_\mathrm{shock}$, and from there we finally get 
\begin{equation}
\zeta_\mathrm{shock} = f_\mathrm{shock} \mathrm{min}(\Delta x, \Delta y, \Delta z)^2, 
\end{equation}
which is then used as an equation term. 

\section{Sink particle method}
\label{sec:sink}

The sink particle method was inspired by \citet{Lee2014}. However, the method
we implemented is not identical to theirs, for two main reasons. First, some
aspects of \citet{Lee2014} implementation are not GPU/Astaroth DSL friendly.
Secondly, we found that if implemented as described in the paper, the method
would not result in numerically stable results. 

The sink particle effects appear in three terms within the system of equations,
with $S_M$ in Equations (\ref{eq:continuity}) and (\ref{eq:momentum}), and
$\mathbf{g}$ in Equation (\ref{eq:momentum}). The mass accretion is performed
by 
\begin{equation}
    S_M(\lnrho, \Delta t) = \frac{\rho - \rho_\mathrm{sink}}{\Delta t}
\end{equation}
within the sink radius $R_\mathrm{sink} = 2 \Delta x$, if $\rho - \rho_\mathrm{sink} > 0$. 
$\Delta t$ is the length of the time step and
$\rho_\mathrm{sink}$ is the value of Equation (\ref{eq:initrho}) at $r =
2R_\mathrm{sink}$. Due to the numerical method, $\rho = \exp(\lnrho)$. During
the code testing we found that a fixed density roof inside the sink particle
radius, respective of the initial condition as above, would be most beneficial
in terms of the system stability. If, e.g., $\rho_\mathrm{sink} = 0$, this
would cause a huge density discontinuity at the sink particle boundary
resulting in strong numerical instabilities. The $S_M$ term in Equation
(\ref{eq:continuity}) takes care of the loss of momentum by the mass accretion. 

The mass accumulation requires a GPU specific implementation. CUDA kernel
operations are performed in multiple threads, and those threads cannot
communicate with each other during the kernel runtime. We include an accretion
buffer, which collects mass accumulation via each thread. At the end of a time
step, accumulated masses per grid element are computed together by Astaroth
collective diagnostic operations, and that mass is added to the sink particle,
$M_{*}$. 

The mass of the sink particle functions as a central attractor, which
contributes to the gravitational term in Equation (\ref{eq:momentum}),
\begin{equation}
    \mathbf{g}(r) = -\frac{GM_{*}}{r^2}\mathbf{\hat{r}},
\end{equation}
active in the domain where
$R_\mathrm{sink} < r < R_\mathrm{max} = 9300\au$. 
Setting a maximum gravity radius, reduces significantly all problematic effects
caused by the corners of a Cartesian domain. In particular, these effects
include pronounced disturbances on the symmetry of the collapsing cloud and
numerical instability in the domain corners which are greatly reduced when
applying  the $R_\mathrm{max}$ limit. In addition, without the $R_\mathrm{max}$
limit the system corners have a tendency to become numerically unstable.
$R_\mathrm{max}$ is slightly smaller than the maximum possible radius, because
that was deemed beneficial during the testing stage.

It should be noted that we use the central mass as the only source of
gravitational force, i.e., we ignore the self-gravity of the infalling gas. The
initial Bondi radius is $GM_{\rm sink}/c_s^2 = 12700\au$, which is larger than
$R_\mathrm{max}$. Therefore the computational box is enclosed in the
gravitational ``sphere of influence'' of the sink particle and the calculation
is self-consistent. We have built this model as being dominated by a central
mass, because for this proof of concept study, a Poisson equation/self-gravity
solver has been out of the scope. However, in future work, as a self-gravity
solver for Astaroth will develop to a functional stage, self-gravity will be
included into the models.

\bibliographystyle{aasjournal}
\bibliography{references}

\end{document}